\documentclass{aa}  
\usepackage{graphicx}
\usepackage{txfonts}
\usepackage{hyperref}
\usepackage{gensymb}
\usepackage{xcolor}
\newcommand{\xin}[1]{#1}
\usepackage[normalem]{ulem}
\usepackage{array} 
\titlerunning{Dwarf Galaxies in the Fornax-Eridanus Supercluster}
\authorrunning{X. Xu et al.}
\begin{document}

   \title{The influence of the Cosmic Web on the properties of dwarf galaxies in the Fornax-Eridanus Supercluster
   }
   \author{X. Xu\inst{1}, P. Ravichandran\inst{1}, R. F. Peletier\inst{1}, Junais\inst{2}, M. A. Raj\inst{1}, P. Awad\inst{3}, R. Smith\inst{4,5}}
   \institute{Kapteyn Astronomical Institute, University of Groningen, PO Box 800, 9700 AV Groningen, The Netherlands
         \and
             Instituto de Astrofísica de Canarias, Vía Láctea S/N, E-38205 La Laguna, Spain
         \and
             Leiden Observatory, Leiden University, PO Box 9513, NL-2300 RA
Leiden, The Netherlands
         \and
             Departamento de Física, Universidad Técnica Federico Santa María, Avenida Vicuña Mackenna 3939, San Joaquín, Santiago de Chile
         \and
             Millennium Nucleus for Galaxies (MINGAL)}
   \date{}
  \abstract
   {In this paper, we analyze a sample of low surface brightness dwarf galaxies with mean effective surface brightness $\bar{\mu}_{e,g} > 24.2$ mag arcsec$^{-2}$, detected using interpretable machine learning tools from the DES survey, and study their properties as a function of their position within the Cosmic Web, to understand the role that the Cosmic Web plays in the evolution of dwarf galaxies. Given their diffuse nature, they are particularly well suited for examining environmental effects, which can affect their morphology and structure. We use the sample of \citet{Tanog2021}, who identified a large number of dwarfs using machine learning, supplemented by the sample from \citet{Thuruthipilly24}, another ML determined sample. \xin{We note that these samples likely contain a significant percentage of false positives; however, they represent the best sample currently available for this work.} We concentrate on the Fornax-Eridanus Supercluster area, in which recently our group \citep{MAraj} determined the 3D filamentary spine (or central axis) of the Supercluster using massive galaxies in this area.
   }
   {Our objective is to explore the effect of the large-scale environment on dwarf galaxies in the Fornax-Eridanus Complex. To do this, we compare the properties of dwarfs in clusters, groups, and the field, and examine how these properties change as a function of distance to the spine of the Fornax Wall. In particular, we investigate whether dwarfs trace the Fornax Wall spine, which has been defined by massive galaxies (\citealt{MAraj}).}
   {We first identify members of the Fornax Wall from the photometric dwarf catalog by \citet{Tanog2021} and \citet{Thuruthipilly24} and we divide the population into two groups: i) those within one virial radius of a galaxy group or cluster and ii) those outside this radius (field galaxies). \xin{We assume that the dwarf galaxies near the Fornax Wall are at the same distance as the massive galaxies in the wall.} We then study their distribution within the complex. We probe the morphology-density relation and examine galaxy properties such as effective radius, surface brightness, S\'ersic index, colour, and stellar mass with respect to their distance from the spine of the Fornax Wall.}
   {Our findings show that red dwarfs are located mostly in and near groups close to the Fornax Wall, where they dominate the population, while blue dwarfs dominate in the field environment. We find that the larger-sized red dwarf galaxies tend to reside in group environments, with significantly larger effective radii compared to their counterparts in the field. Furthermore, red dwarfs are more concentrated towards the Fornax Wall than blue dwarfs. This suggests that the group environment plays a significant role in the evolution of dwarf galaxies. We find that the mass density distribution in field and group/cluster is similar,indicating that the group/cluster population could be an aged version of the field. The group/cluster objects with excess sizes must have been made through interactions in the groups/clusters. This work lays the foundation for future studies that explore in detail the importance of local and large-scale environments in the formation and evolution of dwarfs.}
   {}

   \maketitle
%

\section{Introduction}
\label{sec:intro}
On large scales, the distribution of matter in the Universe forms a complex, web-like network known as the Cosmic Web (e.g., \citealt{Zeldovich1970,deLapparent1986,Bond1996}). This structure consists of interconnected filaments, walls, dense clusters, and vast underdense voids, emerging from the gravitational anisotropic collapse of primordial density fluctuations (\citealt{Zeldovich1970,Peebles1980}). \xin{Matter flows from underdense regions into overdense ones, giving rise to a wide range of environments that provide distinct conditions for galaxy formation and evolution.} At the intersections of filaments, clusters of galaxies-the largest gravitationally bound structures in the Universe-form, while filaments, walls, and voids \xin{trace progressively lower-density regimes}. Groups, clusters, and filaments can \xin{themselves} be embedded within larger systems known as superclusters (\citealt{deVaucouleurs1958-Supercluster}), \xin{which may in turn reside within extended wall-like structures} (\citealt{Einasto-wall-Supercluster,Einasto2017}).

Galaxies are the fundamental building blocks of cosmic structure, and their formation and evolution are closely linked to their location within the Cosmic Web. The properties of galaxies, such as star formation rate, stellar mass, colour, morphology, stellar age, metallicity and $[\alpha/\mathrm{Fe}]$ abundance ratio show clear correlations with the large-scale structure (e.g.,\citealt{Kraljic2020,Winkel2021,Darvish2024, Zarattini2025}). For instance, galaxies in filaments tend to have lower star-forming activity and favour more early-type morphologies compared to field galaxies (\citealt{O'Kane2024}). Similarly, galaxies located within clusters are more often early-type, passive galaxies compared to those in the field, reflecting an accelerating evolutionary process (\citealt{Einasto2007a,Einasto2007b, Lietzen2012}). The environment within superclusters is markedly different from less dense regions, as galaxies in these regions experience stronger gravitational interactions, higher densities of hot gas, and enhanced star formation quenching mechanisms (\citealt{Aghanim2024}). \xin{As such, superclusters and their filamentary connections provide natural laboratories for studying environmental effects on galaxy evolution} (\citealt{Alpaslan2016,Sinigaglia2024}).

\xin{Among galaxies, dwarf systems are especially sensitive to their surroundings.} Dwarf galaxies, the most numerous type of galaxy in the Universe, are characterised by low stellar masses ($\lesssim 10^9\,M_\odot$; \citealt{Bullock2017}) \xin{and shallow gravitational potential wells, making them highly susceptible to environmental influences (e.g., \citealt{El-Badry2016, Martin2019, Boselli2022}.} Therefore, they serve as ideal probes for studying various processes on galaxy evolution in large-scale environments. They \xin{are commonly} classified according to their star formation activity into quiescent, early-type dwarfs (dEs) and gas-rich, star-forming, late-type dwarfs (dIrrs). \xin{Observations} indicate that dwarf galaxies follow a morphology–density relation similar to that observed in massive galaxies (\citealt{Dressler1980}), which found that early-type ellipticals and lenticulars are more abundant in clusters than late-type spirals and irregulars. dEs are predominantly found in high-density environments such as galaxy clusters, whereas dIrrs tend to reside in lower-density regions of the cosmic web or in the field (e.g., \citealt{Bingelli1987-Virgo, Geha2012, Ricciardelli2017, Venhola2019}). Environmental mechanisms, including ram pressure stripping (\citealt{Junais2022}), tidal interactions (\citealt{Gnedin2003}), and galaxy harassment (\citealt{Moore1996harassment-Environment}), play a significant role in shaping the properties of dwarf galaxies.

\xin{Low-surface-brightness galaxies (LSBGs) form an important subset of the dwarf galaxy population. These systems are defined either by the}  central surface brightness \xin{being} fainter than the night sky \citep{BothunImpey1997-LSBdef} or \xin{by the} mean effective surface brightness \xin{that is} fainter than 23~mag~arcsec$^{-2}$ in the $r$-band \citep{Venhola2021}. Observations suggest that faint LSB dwarf galaxies in galaxy groups and clusters can have remarkably similar structural properties \xin{despite residing in different} large-scale environments. For example, \citet{Venhola2017} find that LSB dwarf galaxies in two different environments - the Centaurus group and the Fornax cluster - share comparable colours, sizes, and Sérsic indices, especially at the faint end, \xin{while brighter dwarfs in the group environment tend to be bluer and more compact, likely reflecting weaker environmental processing.} Moreover, the relatively low galaxy velocity dispersions in groups enhance the probability of strong tidal interactions during close encounters. Such tidal effects, possibly amplified by dwarf–dwarf interactions, have been proposed as one of the formation mechanisms for Ultra Diffuse Galaxies (UDGs), particularly in low-mass environments.

\xin{Environmental processing may begin before galaxies enter the dense cores of clusters, through processes commonly described as pre-processing, and filamentary environments are thought to play a key role in this early evolution} (\citealt{Fujita2004,Sarron2019,casti22,Castignani22,Stephenson2025}). For instance, \citet{Chung2021} find that star-forming dwarf galaxies (SFDGs) in the filaments surrounding the Virgo cluster often show lower gas-phase metallicities and higher specific star formation rates (sSFRs) compared to their cluster counterparts. However, variations exist between filaments; in particular, the Virgo III filament shows properties more closely aligned with those in the cluster, including suppressed star formation and enhanced metallicity. Similar trends are also observed by \citet{casti22}, who find that galaxies in filaments already show declining gas content and increasing quenching fractions, especially in high-density regions near filament spines. They further quantify this environmental influence by analyzing the quenching fraction ($f_{\mathrm{Q}}$) as a function of distance to the filament spine. The quiescent fraction decreases from $\sim 0.6$ in the central regions to $\sim 0.2$ at $\sim 1\ h^{-1}\ \mathrm{Mpc}$, indicating that filaments begin to affect star formation primarily within this scale. The quenching is particularly significant for early-type galaxies (ETGs), which maintain high $f_{\mathrm{Q}}$ at all distances, whereas late-type galaxies (LTGs) show a clear decline with increasing separation. These trends suggest that local interactions within filaments can drive early evolutionary transformations, with some dwarf galaxies already showing transitional morphologies prior to cluster infall.

\xin{A nearby laboratory well suited for studying these processes is }the Fornax-Eridanus Complex, a prominent structure consisting of the Fornax cluster, its infalling group, the Eridanus supergroup, the Dorado group, and a number of smaller merging systems (\citealt{Nasonova2011, Makarov2011}). These systems are embedded within a flattened, wall-like large-scale structure known as the Fornax Wall (\citealt{Fairall94}), which itself forms part of the Southern Supercluster Strand - a filamentary extension of the Laniakea Supercluster \citep{Tully2014}. Recent work by \citet{MAraj} (hereafter R24) applied a filament detection method based on machine learning to map the filamentary network \xin{in this region}, identifying 27 filaments and examining how galaxy morphology of massive galaxies varies with filament environments. R24 further suggest that the fraction of early-type and late-type massive galaxies in groups varies systematically with distance to the filament spine (central axis), whereas no such gradients are observed for galaxies outside groups. This supports a role for large-scale structure in shaping the morphology of massive galaxies. \xin{Given their enhanced sensitivity to environmental effects, dwarf galaxies} may show even stronger dependencies on large-scale environment \xin{than massive systems} \citep{Binggeli1988}. While only a few studies have explored the properties of dwarf galaxies within supercluster environments (e.g., \citealt{Mahajan2010, Mahajan2011, Zanatta2024}), these works suggest that morphology and star formation activity correlate with local environmental conditions.

\xin{In this study, we investigate how the properties of dwarf galaxies depend on their location within the large-scale environment of the Fornax-Eridanus Complex. We consider a wide area covering clusters, groups, filaments, and field regions, extending previous R24 work to a homogeneous and statistically significant sample of faint systems.} \xin{A major challenge in studying dwarf galaxies beyond the Local Group is the lack of direct distance measurements. For such low-surface-brightness systems, spectroscopic redshifts are difficult to obtain for large samples: emission-line redshifts are available only for the small fraction of strongly star-forming dwarfs, while quiescent dwarfs require extremely long integration times even on 8 m class telescopes due to their faint surface brightness. In our sample, a cross-match with DESI DR1 provides redshifts for only 19 objects (<0.1\%), showing the severity of this limitation, although only a small fraction of our sample overlaps with the DESI footprint. Alternative distance indicators, such as TRGB \citep[e.g.][]{Vansevicius2025} or surface-brightness fluctuations \citep[e.g.][]{Greco2021,Foster2024}, are limited to distances of a few to $\sim 10\ \mathrm{Mpc}$ \citep{Carlsten2022}, and therefore cannot be applied to most of our sample. As a result, distance measurements are available for only a very small fraction of dwarfs beyond the Local Volume, making a direct three-dimensional environmental classification impossible for large samples. Therefore} we make the critical assumption that all galaxies in our sample reside within the Local Cosmic Web at the distance of the supercluster. \xin{This approach is commonly used in studies of faint dwarf populations \citep[e.g.][]{Paudel2023}.} To minimise contamination from foreground and background objects, we apply multiple membership-selection methods, as detailed in Section \ref{sec:methods}. Under this assumption, the sample LSB galaxies fall within the absolute magnitude range typical of dwarf galaxies\xin{, and we therefore refer to them as dwarfs throughout this paper.} 

\xin{Our dwarf galaxy sample is drawn from the catalogue of} \citet{Tanog2021} (hereafter T21) and its recent extension by \citet{Thuruthipilly24} (hereafter T24), both are based on imaging from the Dark Energy Survey (DES). \xin{The combined catalogue is} selected from a wide-area DES dataset \xin{using} a surface brightness selection criterion of mean $g$-band surface brightness within an effective radius fainter than 24.2 mag arcsec$^{-2}$. T21 compiled the largest LSBG catalogue to date and found that red LSBGs are more strongly clustered than blue ones, independent of stellar mass. This suggests that the observed segregation is driven by environmental factors rather than intrinsic properties. T24 recently expanded the original T21 catalogue by applying advanced transformer-based models to DES data, identifying additional LSBGs \xin{and improving sample completeness}. \xin{Our results are dependent on the reliability of the adopted T21+T24 sample.} \xin{By combining these datasets and using the filamentary structure identified by R24,} we explore how these dwarf galaxies show similar or stronger environmental dependencies, which remain less studied in supercluster environments. Specifically, we analyze how these galaxies are distributed relative to the Fornax Wall and examine whether their morphological properties follow the morphology–density relation on large scales. This work contributes to a broader understanding of the evolution of faint galaxies within extremely dense cosmic environments, such as galaxy superclusters and complexes of the large-scale structure.

This paper is organized as follows. In Section \ref{sec:data}, we describe the observational data and the dwarf catalog used in this study. Section \ref{sec:methods} outlines the methods for identifying supercluster dwarf galaxies and calculating their properties. In Section \ref{sec:results}, we present the results, including the morphological properties and distribution of dwarfs relative to the Fornax Wall. Section \ref{sec:discussion} discusses the environmental impact on dwarf galaxies, focusing on the differences between group and field environments. Finally, we summarise our findings in Section \ref{sec:conclusion}.
\section{Observational data}
\label{sec:data}
In order to perform a statistical study on the dwarfs of the Fornax Wall, a large, homogeneous and highly complete sample is required. Given the very low surface brightness of dwarf galaxies, their distance determination beyond $\sim 5$ Mpc is very difficult, so that additional data such as HI measurements (e.g., \citealt{casti22,Castignani22}) is required, although this method is biased towards late-type galaxies. However, since HI is scarce in clusters, more elaborate techniques are necessary when studying dwarf samples across a wide range of environments. One possibility is the use of surface brightness fluctuations \citep{Carlsten2022}. Alternatively, one can use machine learning, as in T21 and T24, to identify dwarf galaxy candidates from imaging data. While this does not provide accurate distances, we assume that galaxies projected near the spine lie at its distance.

The Dark Energy Survey (DES) uses the ground-based Dark Energy Camera (DECam; \citealt{DEcam}) to obtain deep imaging in the optical SDSS $g, r, i, z$ bands in the Southern Hemisphere. The region used in this study covers $\sim 1750$ square degrees and fully includes all the Fornax Wall groups. T21 provides a dwarf catalogue based on DES. We refer to T21 for its completeness. \xin{The performance of the machine-learning classification in T21 was evaluated using an independent, visually inspected test set designed to distinguish genuine low-surface-brightness galaxies from artifacts. For their DeepShadows convolutional neural network, \citet{Tanoglidis2021} report a completeness of 94.4\% and a purity of 90.3\%, with corresponding 95\% confidence intervals of 0.935–0.953 and 0.891–0.914, respectively.}  In the Fornax Cluster, the T21 catalogue contains $\sim$\;45\% of the dwarf galaxies in the complete \xin{Fornax Deep Survey dwarf} catalogue (FDSDC; \citealt{Venhola2018}). A comparison of their magnitude and stellar mass distributions is shown in Appendix~\ref{appendix-a}. \xin{To further assess the completeness of the DES-based catalogue, we performed mock galaxy injection and recovery tests on DES tiles; the resulting completeness as a function of surface brightness and stellar mass is presented in Appendix \ref{appendix-d}.}

The catalogues used in T21 are based on imaging data from the Dark Energy Survey Data Release 1 (DES DR1; \citealt{Abbott2018}) and the DES Y3 Gold catalog (\citealt{Sevilla-Noarbe21}). The median 3$\sigma$ surface brightness limits are approximately 28.3, 27.9, and 27.4 mag arcsec$^{-2}$ in the $g$, $r$, and $i$ bands, respectively \citep{Tanoglidis2021}. The Y3 Gold catalogue has median coadded magnitude limits of about $g \approx 24.3$ mag, $r \approx 24.0$ mag, and $i \approx 23.3$ mag at a signal-to-noise ratio of 10 \citep{Sevilla-Noarbe21}.

The methods used by T21 to produce the dwarf catalogue are summarised briefly here. They first removed artifacts from the Y3 GOLD catalog  using colours, effective radius, ellipticities and surface brightness obtained by {\ttfamily SourceExtractor} (\citealt{Bertin2010}). They applied specific limits on colours, restricting objects to the range $-0.1 < g-i < 1.4$, and using the conditions 
\[
(g-r) > 0.7 \times (g-i) - 0.4 \quad \text{and} \quad (g-r) < 0.7 \times (g-i) + 0.4,
\]
as well as requiring ellipticity $< 0.7$. The definition of dwarfs adopted was from \citet{Greco2018} as $R_{e,g} > 2.5^{\prime\prime}$ and $\bar{\mu}_e (g) > 24.2$ mag arcsec$^{-2}$. Additionally, a support vector machine (SVM) classification algorithm was used to remove contaminants from diffraction patterns, bright regions of Galactic cirrus, substructures of large spirals and tidal ejecta of high surface brightness galaxies. After visual inspection of these galaxies, \texttt{GalfitM} \citep{2013MNRAS.430..330H} was used to fit a single-component S\'ersic function to galaxy radial profiles. 

Some extended dwarfs can also be massive (e.g., \citealt{Bothun1985}). However, due to the surface brightness-luminosity relation (\citealt{Binggeli1984}), the sample does not contain any non-dwarf LSBs in the Fornax–Eridanus supercluster. In addition to this, the DES reduction pipeline is optimised for faint and small galaxies which can exclude detection of extended (and massive) LSB objects. 


\subsection{Dwarf catalogue of \citet{Thuruthipilly24}}
\citet{Thuruthipilly24} propose a new method for identifying dwarf galaxies using transformer models, significantly improving both the efficiency and accuracy of dwarf galaxy detection in large-scale survey data. Building on the methods of T21, their study applies transformer-based architectures to DES data, aiming to identify previously missed dwarfs. They develop an ensemble of eight different transformer models, which outperform the SVM method originally used in T21. In a later study, convolutional neural networks (CNNs) are also tested \citep{Tanoglidis2021}, but the transformer models show improved performance over both SVM and CNN methods.

By adopting this advanced methodology, \citet{Thuruthipilly24} recover most of the 23,790 dwarfs detected by T21 and identify an additional 4083 new dwarf galaxy candidates from DES DR1. Their identification pipeline follows several steps: preselection of objects based on Source Extractor parameters, application of the transformer models to the selected sample, Sérsic model fitting, and a final visual inspection. Their work has expanded the catalogue from T21, and we will use their catalogue as an important complement to the existing T21 catalogue, adding new data to the dwarf sample within the Fornax Wall region.
\section{Methods}
\label{sec:methods}
In this section, we first define the environments within the supercluster. Then, we obtain the supercluster dwarf (SC dwarf) catalog from the combined T21 and T24 catalogues by identifying members using scaling relations and removing known groups that are outside the Fornax Wall. Finally, we also calculate the absolute magnitudes and the stellar mass of the galaxies by adopting distances based on the groups in the Fornax Wall.

\subsection{Regions of interest}
In this section, we present a preliminary catalogue of supercluster dwarfs. In Section~\ref{sec:fornax_wall}, we identify galaxies located within the Fornax Wall using an angular selection criterion. In Section~\ref{sec:group_field}, we separate the population into two: group and field galaxies.

\subsubsection{Galaxies within the 2D Fornax Wall}
\label{sec:fornax_wall}
Filaments and walls usually have diameters ranging from 1 to 7 Mpc h$^{-1}$ (\citealt{Cautun2014}). In R24, a radius of $r = 3$ Mpc h$^{-1}$ was chosen for the Fornax Wall. This value was selected because it maximized the contrast between the local galaxy density along the filament spine and the surrounding underdense regions, based on a local neighborhood analysis of the 3D galaxy distribution. Since this selection was based on the 3D locations of massive galaxies, we adopt the same radius for our analysis.

The T21 catalog, along with the newly added T24 sample, together cover a region significantly larger than the Fornax Wall (see Figure \ref{fig:FE_SC}). In this paper, we assume that the dwarf galaxies near the Fornax Wall are at the same distance as the massive galaxies in the wall. This assumption is based on the fact that dwarf galaxies are almost always found in groups or clusters, close to massive galaxies. \xin{To test the validity of this assumption, we perform numerical experiments using the SIMBA cosmological simulation to quantify the impact of projection effects on our group and filament classifications. These tests, described in Appendix \ref{appendix-c}, show that although a small fraction of dwarfs are projected onto unrelated groups along the line of sight, the overwhelming majority still trace the same underlying filamentary structure. This supports the use of a common distance assignment for dwarfs near the Fornax Wall in our analysis.} In practice, we consider the distance of dwarf galaxies within $r = 3$ Mpc $h^{-1}$ to be the same as the closest position in the spine. Since this radius corresponds to about $10^{\circ}$ at the distance of the spine, we adopt this value as the angular selection criterion for dwarf galaxies from T21 and T24 in our analysis.

\subsubsection{Group and field environment}
\label{sec:group_field}
As the properties of galaxies within groups and outside groups are known to be different, the dwarfs are divided into two populations: all galaxies within one projected virial radius of the groups defined in R24, as well as the Fornax Cluster, are classified as "group galaxies", while those outside one virial radius of these groups and the Fornax Cluster are considered "field" galaxies.

The Fornax Wall contains only one structure that is classified as a "cluster", the Fornax cluster. In this text, we refer to all galaxies located within one virial radius of either groups or the Fornax cluster as group galaxies. One should understand that the field environment includes dwarf galaxies that are located in the outskirts of groups, i.e., beyond one virial radius. These galaxies may still be (partially) influenced by the potential well of nearby groups or clusters and are therefore, not fully isolated galaxies.

\begin{figure}[htbp]
    \centering
    \includegraphics[width=1\columnwidth]{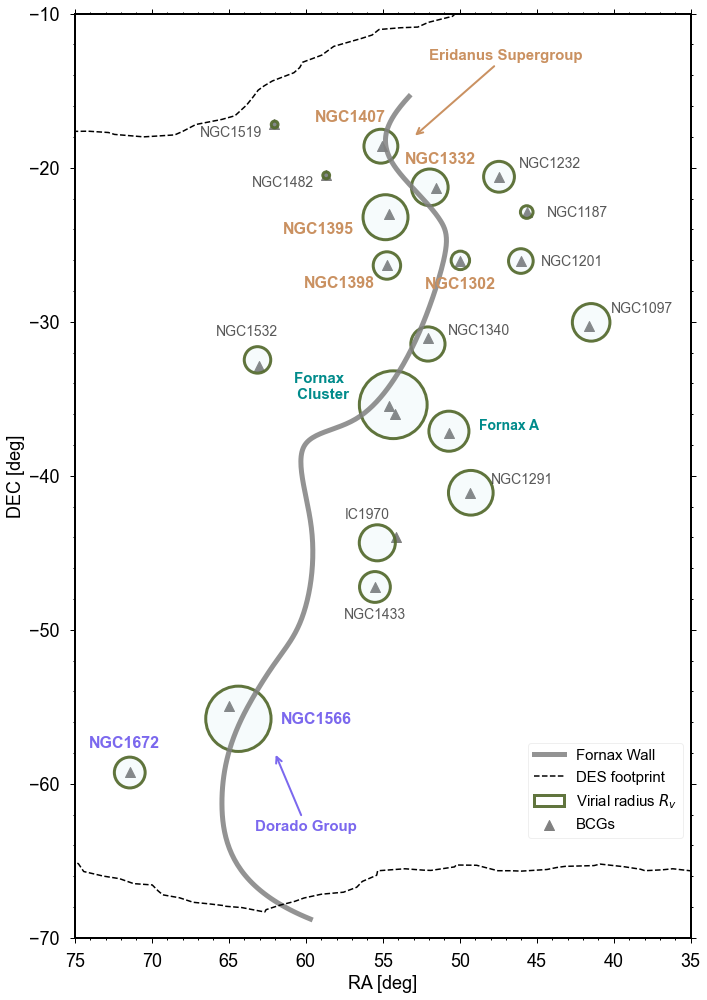}
    \caption{\label{fig:FE_SC} The Fornax Eridanus (FE) Supercluster: The gray line traces the 2D projection of the filaments that form the Fornax Wall spine (R24). Green circles and their corresponding labels represent the Fornax Wall groups (R24, \citealt{Tempel2016-galgroupcat}) and their virial radii (\citealt{Kourkchi2017}). Gray points are the brightest galaxies in clusters and in groups. The Eridanus Supergroup, Fornax cluster and the Dorado group are highlighted with different colours. The dotted-line is the DES footprint of T21 and T24, which encloses nearly all of the Fornax Wall.}
\end{figure}

Figure \ref{fig:FE_SC} illustrates the 20 groups and clusters that belong to the Fornax Wall identified by R24 using the group catalog compiled by \citet{Tempel2016-galgroupcat}. We match the galaxies in the Fornax Wall (from R24) with the galaxy catalogue of \citet{Kourkchi2017} to find their corresponding groups, the group centers and their projected virial radii $R_v$. For the cases with no estimated virial radius, a value of 1 deg is assumed. This is chosen based on the known virial radii of similar small groups, which range from 0.6 to 1.3 deg. The compiled properties of the groups are listed in Table \ref{tab:lsb_hits}.

The galaxies within 10$^\circ$ from the spine are then segregated into group and field galaxies. The catalogue obtained after this step is hereafter called the initial SC (supercluster) dwarf catalogue, which is then used to remove background and foreground galaxies. 

The initial SC dwarf catalogue includes 5240 dwarfs from T21, with 4443 in the field environment and 797 in groups, and 800 additional dwarfs from T24, with 722 in the field and 78 in groups.

\begin{table*} 
\centering
\begin{tabular}{c c c c c c}
\hline\hline
Group ID & RAJ2000 (hms) & DEJ2000 (dms) & $R_{\mathrm{vir}}$ [deg]
 & N$_{dwarf}$ & $\theta$ [deg] \\
\hline
Fornax & 03 37 16.4	& -35 23 22.9	 & 2.2 & 276 & 0.83 \\
Fornax A & 03 22 48.6 & -37 06 30.2	 & 1.3 & 52 & 4.21 \\
NGC1395 & 03 39 16.7 & -23 12 50.8 & 1.46 & 78 & 2.48 \\
NGC1332 & 03 27 46.7 & -21 16 31.8 & 1.19 & 46 & 0.8 \\
NGC1097 & 02 45 52.6 & -30 02 28.0 & 1.22 & 38 & 9.1 \\
NGC1532 & 04 12 31.7 & -32 28 45.5 & 0.86 & 9 & 5.83 \\
NGC1201 & 03 04 07.8 & -26 03 36.0	 & 0.8 & 16 & 4.46 \\
NGC1232 & 03 09 48.2 & -20 35 16.8	 & 1.0 & 12 & 4.51 \\
NGC1302 & 03 19 50.6 & -26 01 03.7	 & 0.6 & 13 & 0.96 \\
NGC1340 & 03 28 15.8 & -31 26 46.3	 & 1.11 & 21 & 0.72 \\
IC1970 & 03 41 24.8 & -44 21 16.9	& 1.17 & 14 & 3.0 \\
NGC1187 & 03 02 36.3 & -22 52 46.9	 & 0.41 & 2 & 5.0 \\
NGC1519 & 04 08 04.4 & -17 11 37.0	 & 0.22 & 0 & 6.92 \\
NGC1407 & 03 40 28.9 & -18 36 15.8 & 1.1 & 106 & 0.31 \\
NGC1398 & 03 38 54.9 & -26 21 00.4 & 0.89 & 28 & 3.12 \\
NGC1482 & 03 54 40.8 & -20 29 40.9	 & 0.22 & 0 & 4.02 \\
NGC1291 & 03 17 07.8 & -41 06 20.2 & 1.45 & 34 & 7.48 \\
NGC1672 & 04 45 42.1 & -59 16 11.6	 & 1.0 & 9 & 3.12 \\
NGC1566 & 04 17 29.2 & -55 46 52.0	 & 2.12 & 112 & 0.11 \\
NGC1433 & 03 42 01.6 & -47 13 19.9	 & 1.0 & 12 & 2.86 \\
\hline
\end{tabular}
\caption{Column 1: Groups within 10$^\circ$ around the Fornax Wall spine (from \citealt{Tempel2016-galgroupcat}). Column 2 and Column 3: Equatorial coordinates of the group center. Column 4: Projected virial radius in degrees from \citet{Kourkchi2017}. Column 5: Number of dwarfs in the initial SC dwarf catalogue (from T21 and T24) within one virial radius. Column 6: Distance of each group from the Fornax Wall.  \label{tab:lsb_hits}}
\end{table*}

\subsection{Membership based on scaling relations}
In this section, we describe the membership selection of dwarf galaxies using two main scaling relations commonly used in previous studies (e.g., \citealt{Venhola2018}):  
\begin{enumerate}
    \item Colour-magnitude diagram (CMD)  
    \item Surface brightness-absolute magnitude diagram  
\end{enumerate}
We use these relations to identify members of the Fornax Wall groups and remove background or foreground galaxies.

\subsubsection{Colour-magnitude diagram}\label{sec:backgroundremovalCM}
Previous studies on dwarf galaxies use scaling relations to remove background galaxies and identify cluster members (e.g., \citealt{Venhola2018}). We justify the usage of these scaling relations to identify the Fornax Wall group and field galaxies in this section.

In the colour-magnitude (CM) space, early-type bright galaxies in clusters form a linear relationship with a nearly constant, negative slope called the red-sequence. 
This sequence reflects a population dominated by old, passively evolving galaxies. Background galaxies, which are redshifted, can appear either below the red sequence if they are very blue, or above it if they are less luminous. A large number of these background bright galaxies were already removed by T21 and T24 due to their selection criteria mentioned in Section \ref{sec:data}. In addition to this, they also explicitly removed galaxies with $g-i > 1.4$. 

Only a few percent of the low-luminosity cluster members are known to lie above the red sequence as outliers (\citealt{Eisenhardt2007}). Compact dwarfs that were formed from larger galaxies are a possible outlier (e.g., \citealt{Price2009}, \citealt{Hamraz2019}). As they are thought to be formed by stripping of the outer parts of a galaxy, generally, they have $R_e(B) <$ 0.2 kpc and surface brightness higher than LSBs (\citealt{Price2009}). The selections made by T21 and T24 have already excluded these galaxies. In addition to compact dwarfs, one might consider whether dusty dwarf galaxies could appear above the red sequence as outliers. However, we argue that this is unlikely. In general, dwarfs are considered to have low amounts of dust as compared to brighter galaxies (see \citealt{Hinz2007}) due to their diffuse nature, their low metallicity and relatively low star formation (\citealt{Vallenari2005}). However, there is some evidence that the amount of dust can vary, and can sometimes be considerable (\citealt{Liang2010}, \citealt{Michea22}, \citealt{Junais2023}). For example, early-type galaxies with blue centers have been observed in far-infrared wavelengths (\citealt{DeLooze2010}). Although dust reddening is usually small for dEs in clusters (\citealt{Hamraz2019}), we do not have sufficient information on isolated dwarfs. Although it is possible that star-forming dwarfs found in isolated regions have dust content that makes them appear red, in practice this is unlikely. This is because their low metallicities limit the amount of dust extinction, and no dwarf galaxies have been reported in the literature where dust reddening causes significant colour shifts. 


\begin{figure}[htbp]
    \centering
    \begin{minipage}{0.5\textwidth}
        \centering
        \includegraphics[width=\textwidth]{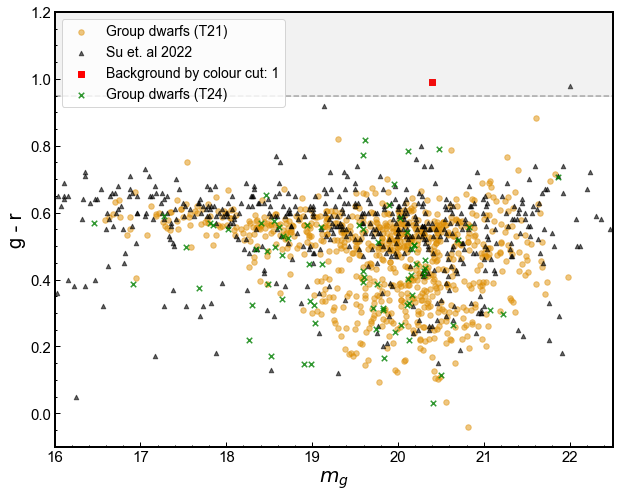}
    \end{minipage}
    \begin{minipage}{0.5\textwidth}
        \centering
        \includegraphics[width=\textwidth]{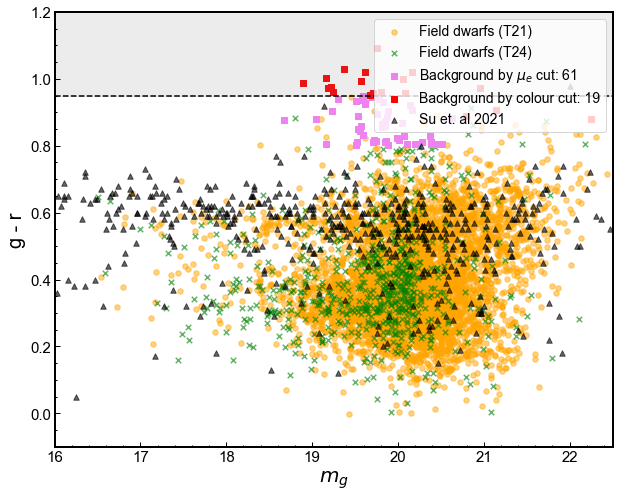}
    \end{minipage}
    \caption{colour-magnitude diagram of group dwarfs (top; orange for T21 and green for T24) and field dwarfs (bottom; orange for T21 and green for T24). Background galaxies removed using colour selection (red), using surface brightness-magnitude diagram (pink; see Section 3.1.4) are included. The gray shaded region is the region removed by the $g - r$ colour selection. Fornax cluster dwarf galaxies from \citet{Su2021} are plotted to highlight the red-sequence (black).}
    \label{fig:CM}
\end{figure}

In this study, we assume that the red sequence is the same for all groups, since all galaxies are situated within 30 Mpc. Following \citet{Venhola2018}, who define Fornax cluster members as $g - r \leq 0.95$ and $g - i \leq 1.35$ based on NGC1399 galaxy, we removed galaxies outside these limits to exclude remaining outliers (see Figure \ref{fig:CM}). To exclude foreground contamination, 19 field and 2 group galaxies from T21, as well as 5 field and 1 group galaxies from T24 with $g - r \leq 0$ and $g - i \leq 0$, are also removed based on the range of colours in previous works on the Fornax cluster (\citealt{Su2021},\citealt{Venhola2021}).
 
\subsubsection{Surface brightness-absolute magnitude diagram} \label{sec:backgroundremovalSB}

\begin{figure}[htbp]
   \centering
   \includegraphics[width=1\columnwidth]{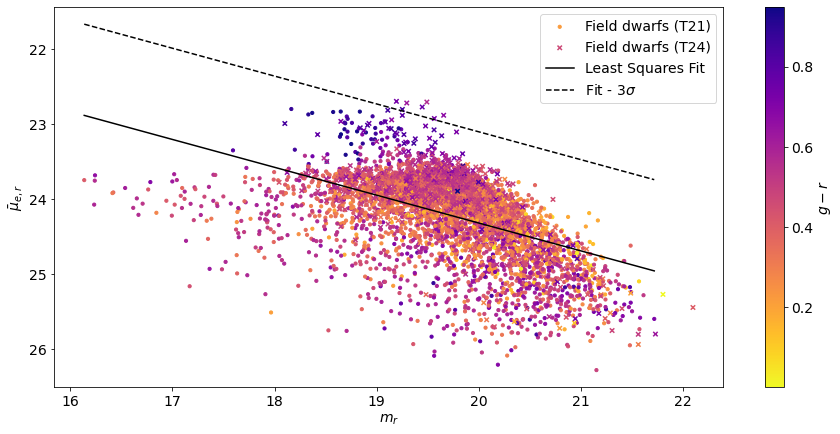}   
    \caption{\label{fig:SurfPrist} 
    Surface brightness-magnitude diagram of field dwarfs (T21: dots, T24: crosses) with $g - r$ colour in the colourbar. The linear least squares fit of the diagram is included as the black line. The dotted line is 3$\sigma$ away from the fit towards high surface brightness. }
\end{figure}

The measured surface brightness of a nearby galaxy does not change with distance because both the flux and the angular size follow the inverse-square law. Therefore, massive background galaxies appear as outliers on the surface brightness-magnitude diagram with fainter magnitudes and higher surface brightness. As done in \citet{Venhola2018}, the sigma-clipped group and field surface brightness-magnitude relation is fit with a linear relation (Figure \ref{fig:SurfPrist}). Galaxies that are $3 \sigma$ from the fit towards higher surface brightness would be considered background. While T21 and T24 both apply a cut of $\bar{\mu}_{\text{eff}}(g) > 24.2$ mag arcsec$^{-2}$ to exclude high surface brightness galaxies, we identify 3 outliers in the T24 sample that lie slightly above this line. In the field environment, a clump of galaxies that were in the range of $0.8 < g - r < 0.9$ mag with $m < 19.5$ mag and $\bar{\mu}_{e} < 23.3$ mag arcsec$^{-2}$ was found as outliers from the general trend (Figure \ref{fig:SurfPrist}).  After visual inspection of this clump, 30 field galaxies were removed as they were bright ellipticals and background spirals galaxies.

\subsubsection{Removal of foreground and background groups} \label{sec:backgroundremovalGroups}

After the removal of background galaxies through scaling relations, the 2D distribution of the dwarfs showed clumping that were not part of the Fornax Wall groups listed in Table \ref{tab:lsb_hits}. To investigate this, we overlay the group catalog by \citet{Tempel2016-galgroupcat} and find that some of the clumping coincide with background or foreground groups. \xin{We remove dwarfs that are spatially associated with foreground or background groups from \citet{Tempel2016-galgroupcat}, where all of the following conditions are simultaneously satisfied:}
\begin{itemize}
    \item[1. ]  Comoving distance of groups $D < 15$ Mpc and $30< D < 150$ Mpc.
    \item[2. ]  Group centers are not within 2 times the virial radius of Fornax Wall groups. 
    \item[3. ]  Group centers are not within 1 deg from massive galaxies.
    \item[4. ]  At least 5 dwarfs are present within 0.5 deg from group center.
\end{itemize} 
The Fornax Wall groups have comoving distances within 15 to 30 Mpc. The 150 Mpc limit for the comoving distance in the first condition is used because our catalogue is estimated to contain dwarfs up to $\sim$ 100 Mpc based on the association of density peaks in the dwarf distribution with known clusters. As small groups that are present within the Fornax Wall were not identified by T21, it is possible that the catalog extends farther than 100 Mpc. \xin{Conditions (2) and (3)} are imposed because some groups are found to coincide with massive galaxies (from R24) in the Eridanus supergroup and the surrounding region and the dwarfs near them are most likely supercluster members. \xin{Finally, condition (4) ensures that the selected systems represent genuine dwarf galaxy groups.} We remove 252 dwarfs from T21 and 29 dwarfs from T24 based on these group membership criteria.

\begin{figure}[htbp]
   \centering
   \includegraphics[width=1\columnwidth]{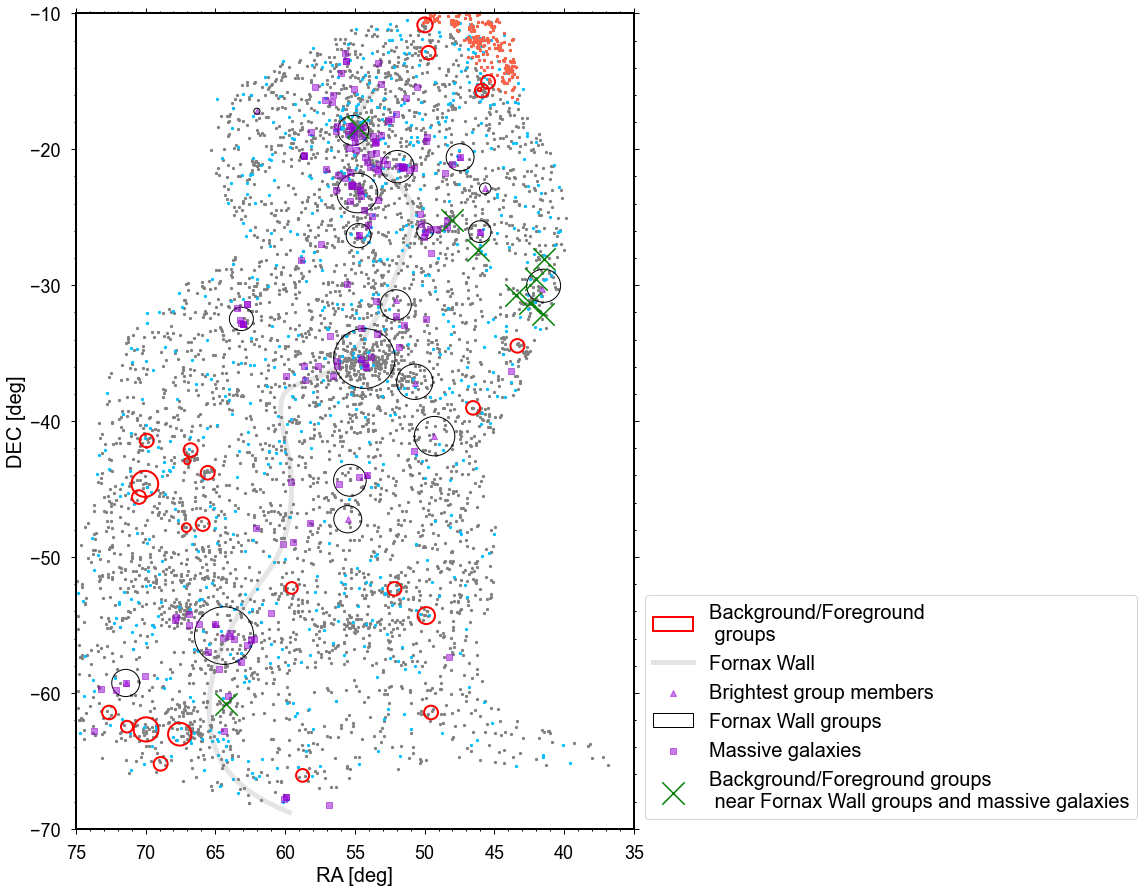}
    \caption{\label{fig:BGgroups} Virial radii of the groups in the Fornax-Eridanus Supercluster region, including Fornax Wall groups (black circle), background and foreground groups with distances outside the 15 - 30 Mpc h$^{-1}$ range (red circle). The background and foreground groups within 2 times the virial radius of Fornax Wall groups and 1 deg around massive galaxies (green cross). Also shown are objects removed manually (red points), brightest group members in the Fornax Wall from R24 (purple points) and dwarfs from the T21 catalogue after background galaxy removal through scaling relation (gray points), and dwarfs from the T24 catalogue (blue points) are also included.}
\end{figure}

Figure \ref{fig:BGgroups} shows the background and foreground groups that pass the above conditions. In total, we remove 237 field dwarfs from T21 and 29 field dwarfs from T24 within one virial radius of these groups. In addition to this, 215 T21 galaxies (red points in Figure \ref{fig:BGgroups}) and 35 T24 galaxies in $55^\circ < $ RA $ < 75^\circ$ and $-20^\circ < $ DEC $ < -10^\circ$ are removed manually after visually inspecting the crowded region. These galaxies show unusually red colors and are clustered but are not part of the Fornax Wall groups. This region lies near the edge of the DES footprint and may be affected by edge effects. It may also contain background clusters at larger distances that are not identified as part of the Fornax Wall. The exclusion of these galaxies does not significantly affect the analysis of field galaxies, as their numbers here are small compared to the total sample. The final SC dwarf catalog contains 3926 dwarfs in the field region and 791 group dwarfs from T21, combined with 640 field and 77 group dwarfs from T24. Group galaxies are those within one projected virial radius of the groups; others are field galaxies. In total, the sample includes 4566 field dwarfs and 868 group dwarfs, which we use for the following analysis.

\subsection{Distances, Absolute Magnitudes and Stellar Mass} \label{sec:Distances}

In this section, we calculate the stellar masses and the absolute magnitudes of the galaxies. To calculate the stellar mass $M_*$, we use the empirical relation produced by \citet{Taylor2011}. They use the $g - i$ colour and $M_i$ as follows: 
\begin{equation} \label{eq:stellarmass}
    \text{log}_{10} (\frac{M_*}{M_\odot}) = 1.15 + 0.70\;(g-i) - 0.4\;M_i.
\end{equation}
where $M_i$ is the absolute magnitude in the restframe $i$-band, expressed in the AB system.

\begin{figure}[htbp]
   \centering
   \includegraphics[width=1\columnwidth]{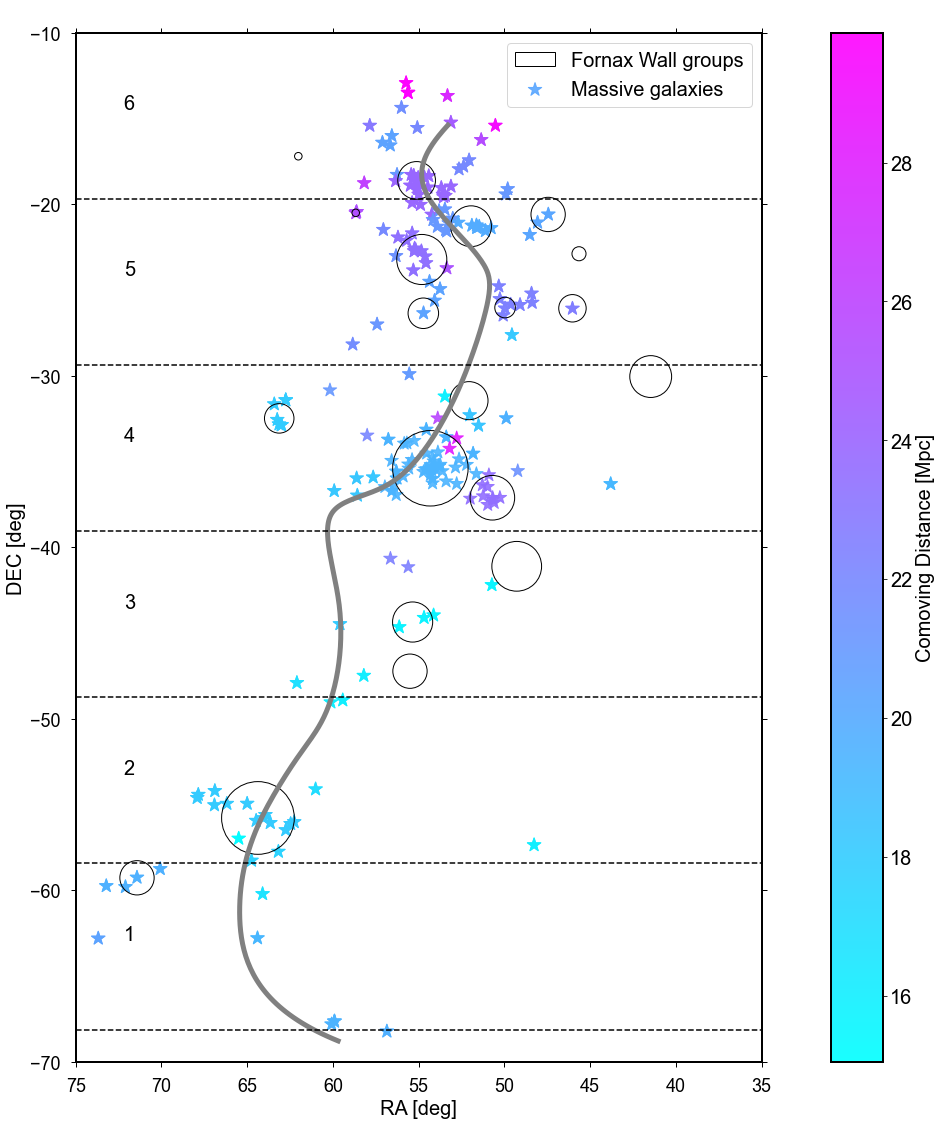}
    \caption{\label{fig:GiantDist} Massive galaxies in the Fornax Wall with their comoving distances from \citet{Tempel2016-galgroupcat} in the colourbar. Black circles are the Fornax Wall groups. Dotted lines separate the Wall into segments which are labelled with numbers.}
\end{figure}

The Fornax Wall has a distance range of 15 - 30 Mpc h$^{-1}$. To determine the distances of the dwarf galaxies, we interpolate across the distances of the massive galaxies. Figure \ref{fig:GiantDist} shows the distances of massive galaxies. They vary smoothly in the bottom half of the Fornax Wall, but vary non-uniformly near the Fornax cluster and the Eridanus Supergroup regions in the 2D space. Using the nearest neighbour to adopt distances can thus introduce an additional bias due to projection effects. Therefore, a more robust approach is required.

In 3D, the Fornax Wall spine is positioned such that the Dorado group region is nearest and the Eridanus Supergroup region is the farthest from us (Figure \ref{fig:GiantDist}). Taking advantage of this, we divide it into six segments based on their declination ranges, as shown in the figure. We adopt the distances of the well-known groups in the segments. Segments 1, 2 and 3 are at a distance of 18.4 Mpc h$^{-1}$ based on the distance of NGC 1566 in the Dorado group. Segment 4 is at a distance of 20 Mpc h$^{-1}$ based on the Fornax cluster. Segments 5 and 6 contain the Eridanus Supergroup and we use the distance of NGC1407 at 24 Mpc h${^-1}$. For the dwarf galaxies, we use the distance of the segment in which they are, and then calculate their absolute magnitude and stellar mass.

\section{Results}
\label{sec:results}
In this section, we first discuss the properties of dwarfs in the SC dwarf catalog obtained in Section \ref{sec:methods}. We investigate the morphology of the sample using the size-magnitude diagram, colour-colour diagram and the S\'ersic index-magnitude diagram. We present the calculated absolute magnitudes and stellar mass. We then investigate the distribution of galaxies with respect to the Fornax Wall spine by using 2D density maps and calculating the fraction of galaxies with different properties as a function of distance from the spine. For each of these results, we compare galaxies in the group and field environment. 

\subsection{Properties of the dwarfs in the supercluster}

In this section, we show the properties of the galaxies in the SC dwarf catalog to discuss their morphology, size and stellar mass. 

\subsubsection{Size-magnitude relation} \label{sec:SizeMag}
\begin{figure}[htbp]
   \centering
   \includegraphics[width=1\columnwidth]{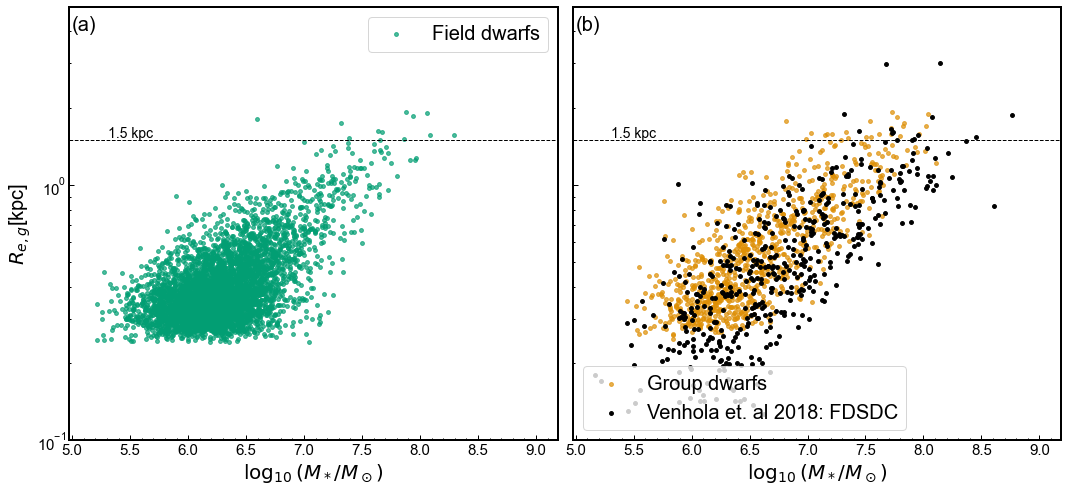} 
    \caption{Size($R_{e,g}$)-stellar mass ($\log M_\star$) diagram for field dwarfs (left) and group dwarfs (right). The dwarf galaxies from FDSDC by \citet{Venhola2018} are also shown with the group dwarfs on the right, in black. The 1.5 kpc line marks the size criterion commonly used to define UDGs.
    \label{fig:size_mass}}
\end{figure}

The size($R_{e,g}$)-stellar mass relation (Figure \ref{fig:size_mass}) is one of the Kormendy relations, which describe relations between photometric properties of galaxies. This relation was presented by T21 for their full sample to demonstrate that their detected objects are indeed dwarf galaxies. Here, we show this relation separately for group and field galaxies. 

\begin{figure}[htbp]
   \centering
   \includegraphics[width=1\columnwidth]{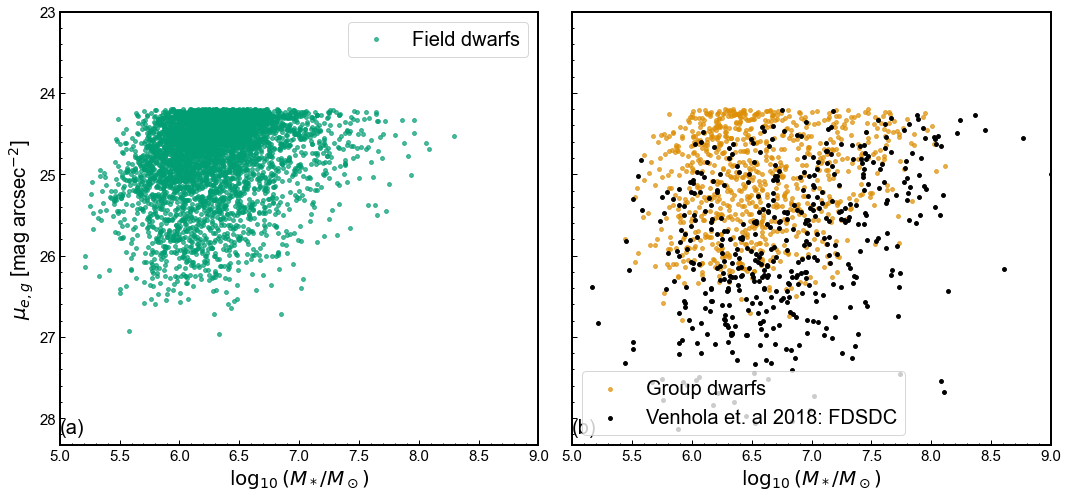} 
    \caption{Same as Fig.~\ref{fig:size_mass} for the surface
brightness ($\mu_{e,g}$)-stellar mass ($\log M_\star$) relation.
    \label{fig:sb_mass}}
\end{figure}

In Figure \ref{fig:size_mass} (b), a number of large dwarfs with $R_{eff} > 1.5$ kpc can be seen. These are referred to as UDGs \citep{vanDokkum2015}. We find 12 UDG candidates in the field and 17 in the group environment. The largest UDGs are found in groups. Comparing field and group dwarfs, we find that large group dwarfs do not have counterparts in the field. This will be further discussed in Section \ref{sec:sizedistribution}. Comparing with the FDSDC sample, we find that group dwarfs are generally slightly more compact than cluster dwarfs, but larger and more diffuse than field dwarfs. Field dwarfs tend to have higher surface brightnesses, while group dwarfs are intermediate, and cluster dwarfs have the lowest surface brightness.

We compare the group dwarfs with the FDSDC dwarf galaxies. Figure \ref{fig:sb_mass} shows that both catalogues reach similar depths, up to $\bar{\mu}_{e,g}$ $\sim$26 mag arcsec$^{-2}$. Some low luminosity dwarfs (with $\log M_\star < 5.5$) are also missing in our catalogue, since the FDSDC adopts a lower limit of $R_{e,g} > 1$ arcsec, while our selection requires $R_{e,g} > 2.5$ arcsec. 

\subsubsection{Colour-colour relation: Defining red and blue galaxies} \label{sec:colourcolour}

\begin{figure}[htbp]
   \centering
   \includegraphics[width=1\columnwidth]{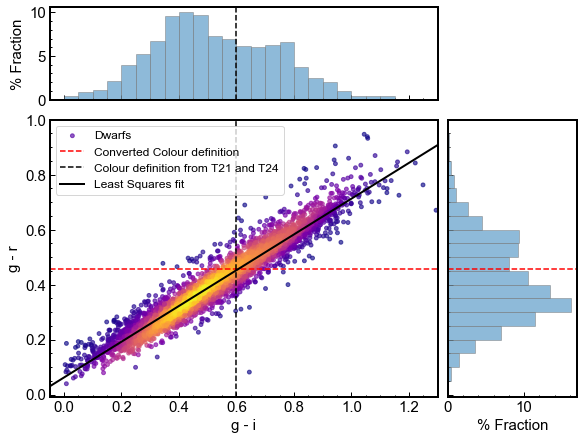} 
    \caption{\label{fig:colorcolor} $g - r$ vs $g - i$ diagram of dwarfs with the density of the scatter plot shown using colourmap. The fit to the relation (black solid line) is included. The definition of galaxy colour adopted by T21 and T24 (black dotted line; $g - i = 0.6$) and the corresponding value used in this paper (red dotted line; $g - r = 0.452$) is included. The histograms show the fraction of galaxies in each axis. }
\end{figure}

In Figure \ref{fig:colorcolor}, we show the colour-colour relation for the dwarfs in the SC dwarf catalog. The histograms for $g - i$ and $g - r$ colours have two peaks, showing the bimodality in the colour distribution. It has been well-established in the literature (e.g., \citealt{Blanton2003}) that red and blue galaxies form two distinct populations. Recent studies (\citealt{Lazar2024}) have shown that this bimodality also holds for dwarf galaxies, with rest-frame ($g - i$) colours exhibiting well-defined red and blue peaks, as reflected in the density distribution in Figure \ref{fig:colorcolor}. In T21 and T24, galaxies with $g - i < 0.6$ were defined as blue (late-type) and those with $g - i \geq 0.6$ were considered red (early-type). This segregation was defined using the intersection point of the Gaussian curves fit around each peak in the $g - i$ histogram. The black dashed line in Figure \ref{fig:colorcolor} represents the colour separation definition ($g - i = 0.6$) used by T21 and T24. We use the same definition as T21 and T24. However, since we use $g - r$, we convert the original colour separation value accordingly. To do this, we fit the colour-colour diagram with a linear relation (slope $=$ 0.65, intercept $=$ 0.06), and obtain a separation value of $g - r = 0.452$.

Figure \ref{fig:colorcolor} shows that blue dwarfs dominate the SC dwarf catalog. Both T21 and T24 demonstrated that their catalog contains more blue dwarfs than red dwarfs, whereas other dwarf catalogs, such as that of \citet{Greco2018}, show equal populations of both. 

\subsubsection{S\'ersic index-magnitude relation} \label{S\'ersicIndex-colour}

\begin{figure}[htbp]
   \centering
   \includegraphics[width=1\columnwidth]{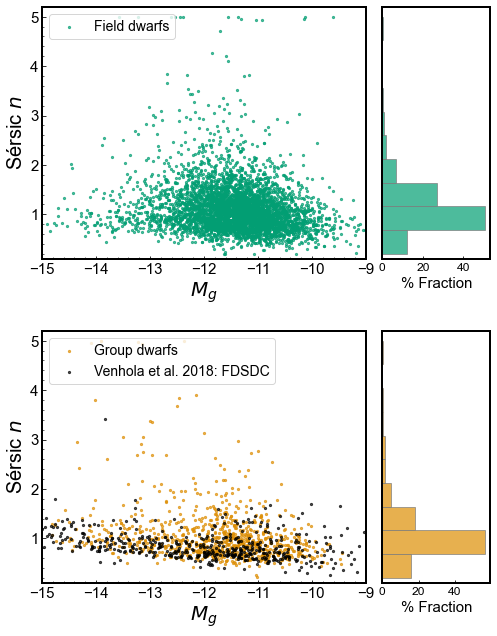} 
    \caption{\label{fig:SersicMag} S\'ersic $n$ vs absolute magnitude $M_g$ diagram of field (top) and group (bottom) dwarfs. The dwarf galaxies from FDSDC by \citet{Venhola2018} is also shown with the group dwarfs (black squares). The histograms show the distribution of S\'ersic $n$ for field (top) and group (bottom) dwarfs.}
\end{figure}

The S\'ersic index $n$ characterises the shape of a galaxy's radial intensity profile. An exponential profile similar to that of a galaxy disk corresponds to $n = 1$. Galaxies with a central concentration and a disk typically have higher $n$. In general, most dwarfs are diffuse and do not have a distinct nucleus. Therefore, they typically have S\'ersic $n \leq 1$ (e.g., \citealt{Venhola2021}). We find that the mean S\'ersic $n$ (measured in the g-band) of the field galaxies is 1.12 and the mean S\'ersic $n$ of group galaxies is 1.07.

Figure \ref{fig:SersicMag} shows the S\'ersic index of field and group galaxies. The dwarf galaxies in FDSDC are mostly limited to $n < 2$ while the SC dwarf catalogue has a large scatter, especially in the field environment. There are several dwarfs with $n > 2$; these could be either red or blue massive background galaxies, or blue dwarfs with a bright nucleus (BCDs). However, it is likely that the photometry of these galaxies is not very accurate, possibly due to the limitations of the single-component S\'ersic fitting used in both T21 and T24. Since the fraction of galaxies in the SC catalogue with n>2 is only 4\%, this will not significantly affect the further analysis.

\subsubsection{Magnitude and stellar mass function} \label{sec:LF}
\begin{figure*}
    \centering
    \begin{minipage}[b]{0.33\textwidth}
        \centering
        \includegraphics[width = \textwidth]{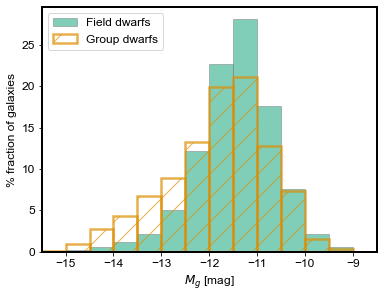}
    \end{minipage}    
    \begin{minipage}[b]{0.33\textwidth}
        \centering
        \includegraphics[width = \textwidth]{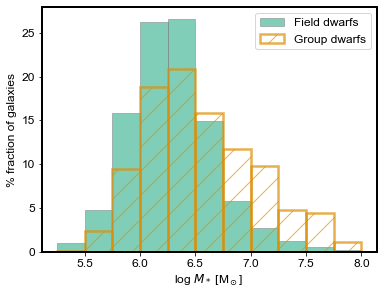}
    \end{minipage}
    \begin{minipage}[b]{0.33\textwidth}
        \centering
        \includegraphics[width = \textwidth]{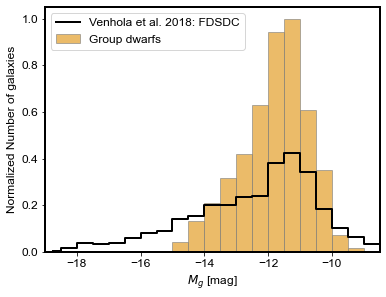}
    \end{minipage}
    \caption{\label{fig:Histograms} The $M_g$ histogram for group galaxies along with the histogram of dwarf galaxies in FDSDC by \citet{Venhola2018} (right) and the histograms of $M_g$ (left), $M_*$ (middle) for the group and field galaxies are shown.}
\end{figure*}

\begin{figure*}
    \centering
    \begin{minipage}[b]{0.33\textwidth}
        \centering
        \includegraphics[width = \textwidth]{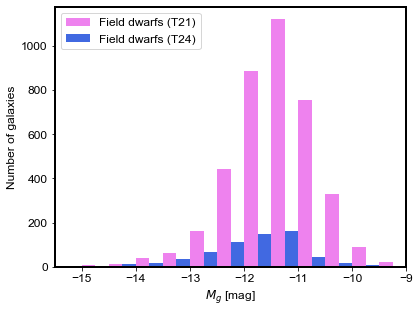}
    \end{minipage}
    \begin{minipage}[b]{0.33\textwidth}
        \centering
        \includegraphics[width = \textwidth]{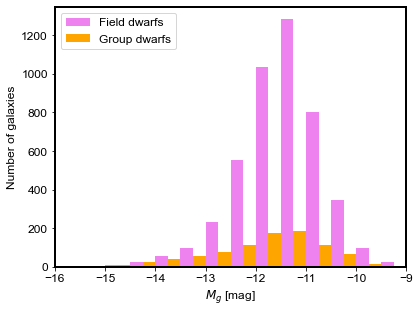}
    \end{minipage}    
    \begin{minipage}[b]{0.33\textwidth}
        \centering
        \includegraphics[width = \textwidth]{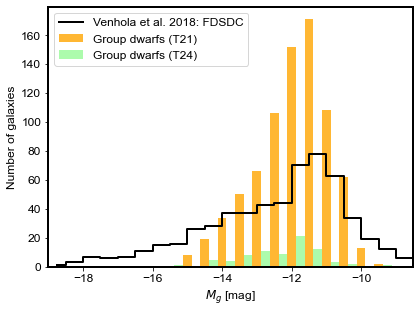}
    \end{minipage}
        \begin{minipage}[b]{0.33\textwidth}
        \centering
        \includegraphics[width = \textwidth]{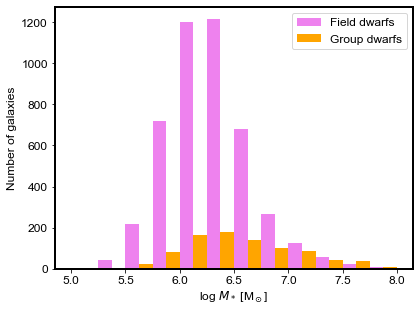}
    \end{minipage}
        \begin{minipage}[b]{0.33\textwidth}
        \centering
        \includegraphics[width = \textwidth]{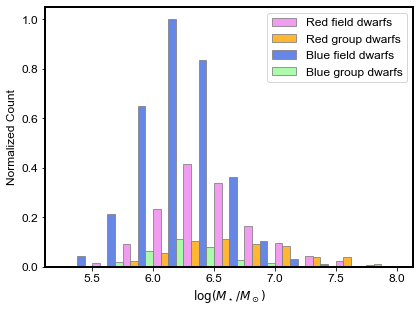}
    \end{minipage}
    \caption{\label{fig:Histograms_t24}
Histograms of $M_g$ and $M_*$ for field and group dwarf galaxies. Top-left: field dwarfs (T21 and T24) in $M_g$. Top-middle: field and group dwarfs in $M_g$. Top-right: group dwarfs (T21 and T24) and FDSDC sample from \citet{Venhola2018} in $M_g$. Bottom-left: field and group dwarfs in $M_*$. Bottom-right: four dwarf galaxy subsamples (red group, red field, blue group, and blue field) in $M_*$. The y-axis shows the number of galaxies per bin.}
\end{figure*}

The histograms of the calculated absolute magnitudes and stellar mass for galaxies in the SC dwarf catalog are shown in Figure \ref{fig:Histograms}. To check if the values agree with the literature, we use the FDSDC galaxies and compare the absolute magnitude $M_g$ in our group catalog. The absence of SC group galaxies between $-16 < M_g < -18$ mag is due to the dwarf selection criteria used in both T21 and T24, namely, that the effective surface brightness in the $g$-band must be fainter than $24.2$ mag arcsec$^{-2}$. According to the effective radius–magnitude relation of Fornax galaxies \citep{Venhola2019}, this corresponds to about $M_r \sim -15$ for red galaxies and $M_r \sim -13$ for blue dwarfs. The range of values in the SC dwarf catalog are within the range in FDSDC. The shape of the histogram is similar in both cases with increasing fraction of galaxies from the bright side to the peak at $M_g \sim -12$ mag after which incompleteness sets in. To further investigate potential differences between the two catalogues, we separately plot these distributions for T21 and T24 in Figure \ref{fig:Histograms_t24}. The histograms of T21 and T24 samples have similar shapes and trends in both environments, confirming that the combined analysis shown in Figure \ref{fig:Histograms} represents the overall properties of the dwarf galaxies from both catalogues. 
Figure \ref{fig:Histograms_t24} also shows the stellar mass distributions of the four subsamples - red group, red field, blue group, and blue field. The red group sample shows the highest stellar mass peak, while the blue group and field samples show similar, lower peaks. This comparison suggests the subtle differences in stellar mass content among these populations.

As a result, Group and field galaxies show different distributions. Nearly 48\% of group galaxies have log $M_* > 6.5 \; M_\odot$ while 25\% of field galaxies have log $M_* > 6.5 \; M_\odot$. 

\subsection{Distribution of galaxies and their structural properties} 
We study the distribution of dwarfs, probe their morphology-density relationship and structure with respect to their distance from the Fornax Wall. 

\subsubsection{The 2D distribution of galaxies} \label{sec:Distribution}

In this section, we describe the distribution of dwarfs within the Fornax Wall. We use Kernel Density Estimation (KDE) in the 2D plane to obtain density contours. Our goal is to identify regions of high density within the supercluster. Therefore, a kernel of size $\sim$ $2 \times 2$ deg$^2$ is used based on the virial radius of the Fornax cluster (2.2 deg). The densities are evaluated on a grid spanning $-70^\circ < DEC < -10^\circ$ along and $35^\circ < RA < 75^\circ$. While KDEs produce a continuous distribution map, the estimated densities can spill over from high density regions to nearby areas. Therefore, it is important to look at the KDE in combination with the 2D distribution of dwarfs, in order to form conclusions. 

\begin{figure*}
    \centering
    \begin{minipage}[b]{0.33\textwidth}
        \centering
        \includegraphics[width = \textwidth]{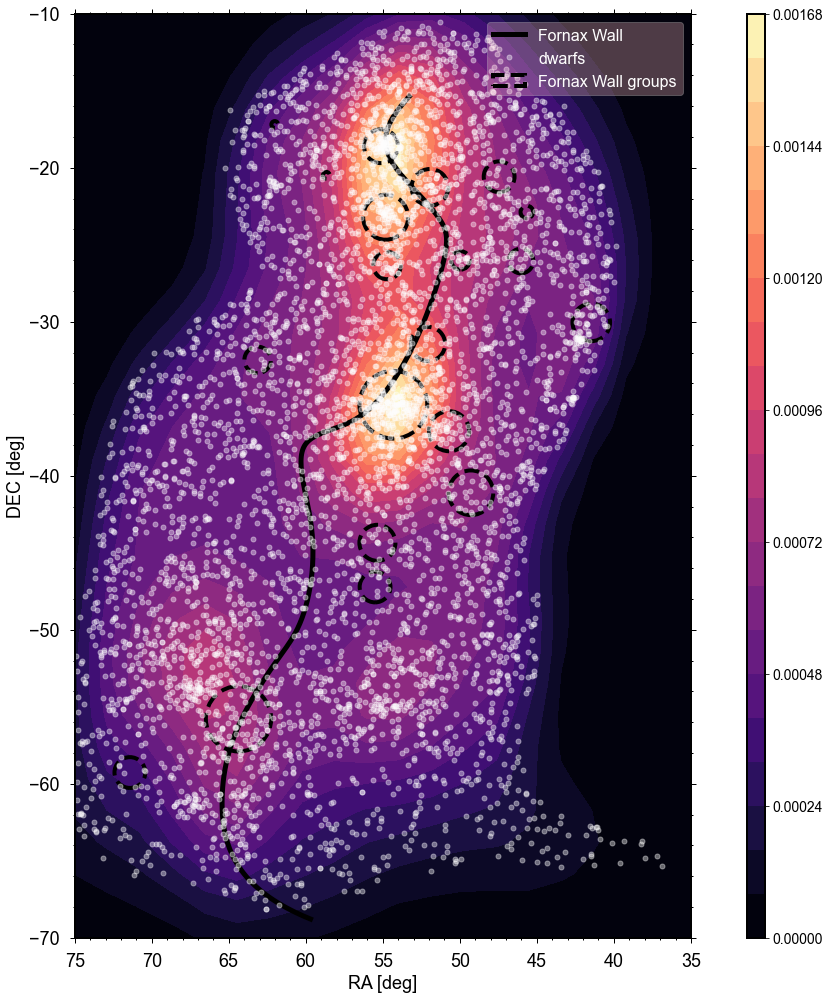}
    \end{minipage}
    \begin{minipage}[b]{0.33\textwidth}
        \centering
        \includegraphics[width = \textwidth]{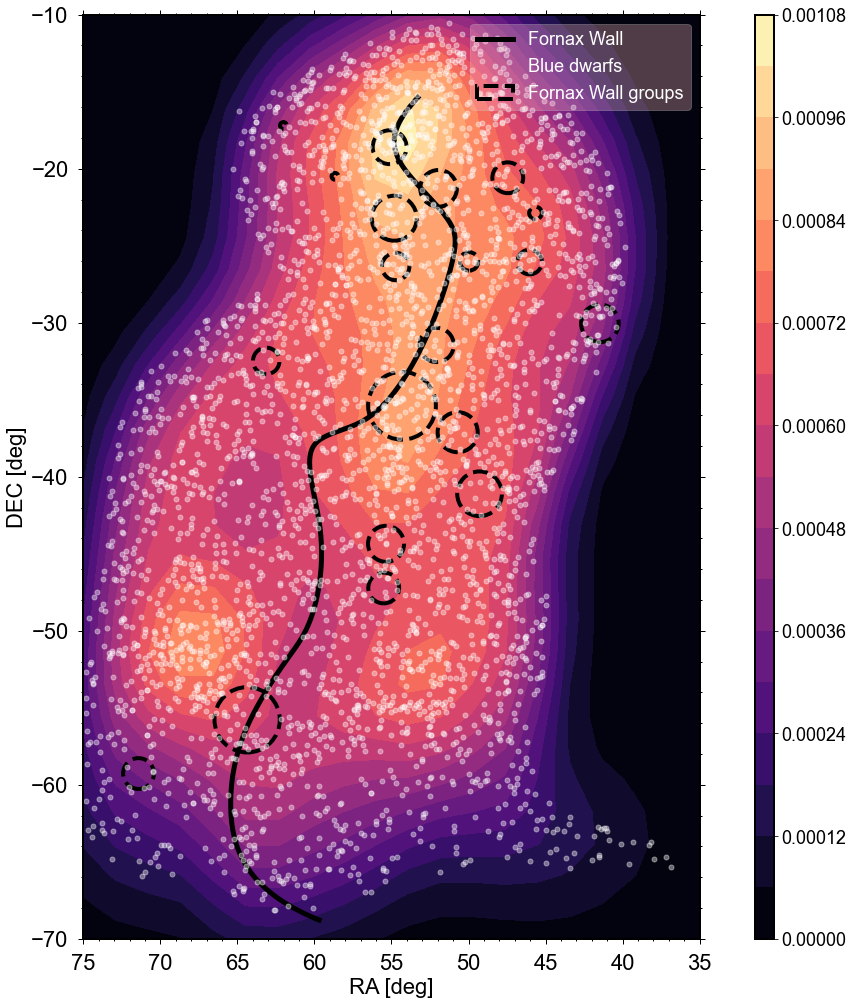}
    \end{minipage}
    \begin{minipage}[b]{0.33\textwidth}
        \centering
        \includegraphics[width = \textwidth]{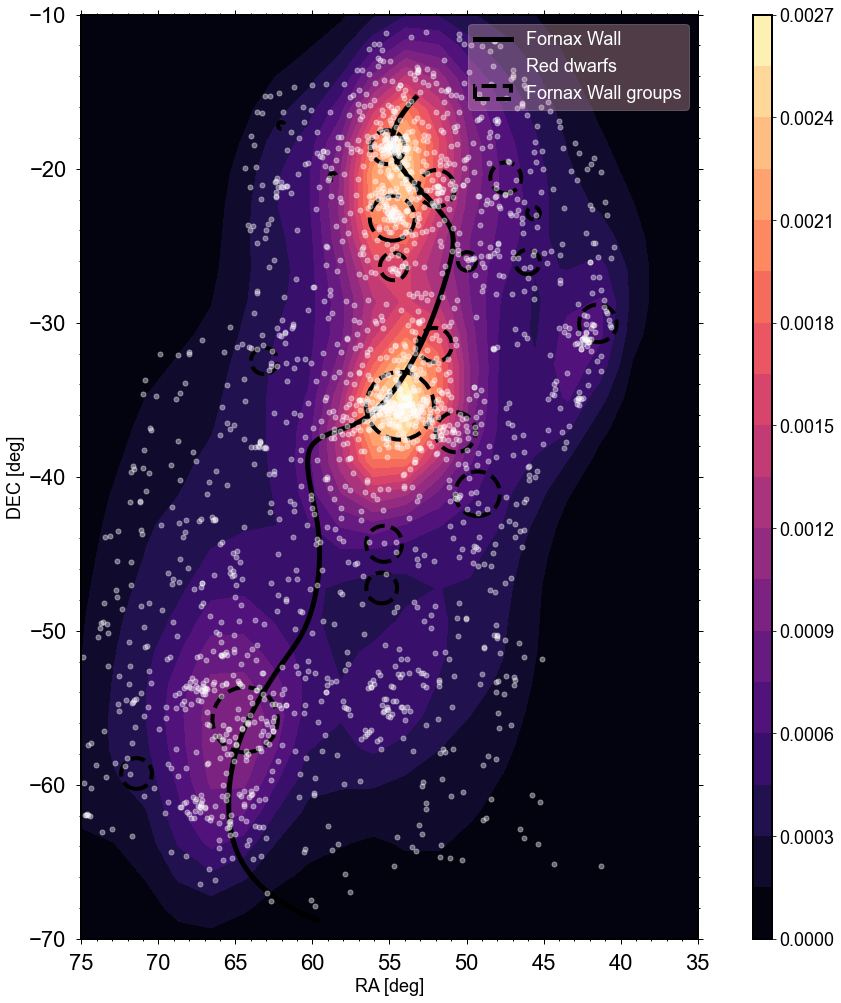}
    \end{minipage}
    \caption{KDE for the spatial distribution of all dwarfs (left), blue dwarfs (middle) and red dwarfs (right) within the Fornax Wall and their 2D distribution (white points). The Fornax Wall (black solid line) spine is also shown alongwith its groups (black dashed-circles). Colorbars show the normalized KDE density of dwarf galaxies.   \label{fig:KDE}}
\end{figure*}

Figure \ref{fig:KDE} shows the distribution of all dwarfs that are part of the Fornax Wall. The galaxies are seen to be clumped in groups closer to the spine, mainly in the Eridanus Supergroup (over-density at the top) and the Fornax cluster region (over-density at the center). We also observe clumping near NGC 1566 of the Dorado Group (over-density at the bottom). Due to the well known morphology-density relation, most of the galaxies in the high density regions are expected to be early-type and red in colour. T21 and T24 both found that red dwarfs were clustered while blue dwarfs were distributed uniformly. To identify if this is the case and if it depends on the proximity to the spine, we show the KDE for blue and red dwarfs separately in Figure \ref{fig:KDE}.

T24 find that blue LSBGs have weaker but noticeable clustering than that of the red population. Our results are consistent: we also detect some concentration of blue dwarfs toward the groups and along the Fornax Wall, although this is much less pronounced than for the red galaxies. In the next section, we will quantify this. Any absence of blue galaxies seen in certain regions (for example, (RA, DEC) $\sim$ (65, -40)) is due to the removal of galaxies within background or foreground groups in Section \ref{sec:backgroundremovalGroups}. In contrast, the red galaxies are clustered in groups closer to the spine and the density of red galaxies is extremely low in the field environment. Red galaxies are, therefore, the main contributors to the gradient observed in the distribution of all the galaxies as expected. 

\subsubsection{Morphology-density relation with respect to the Fornax Wall}
\begin{figure}[htbp]
   \centering
   \includegraphics[width=1\columnwidth]{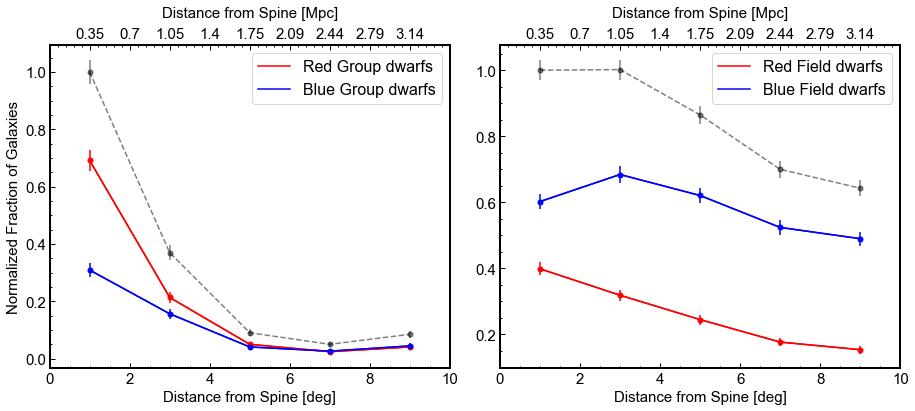} 
    \caption{Fraction of the dwarfs in the SC dwarf catalog that are red/blue in groups (left) and red/blue in the field environment (right) as a function of projected distance from the Fornax Wall spine. Error bars represent the Poisson uncertainties of the counts in each distance bin. The dashed lines represent the fraction of dwarfs that are either red or blue.}
    \label{fig:BlueRedWall}
\end{figure}

In this section, we discuss how the properties of dwarf galaxies vary as a function of their distance from the Fornax Wall spine, with the aim of testing whether the filaments themselves are affecting the evolution and timescales of dwarf galaxies. In Figure \ref{fig:BlueRedWall} we find that for group dwarfs, the fraction of red galaxies decreases steeply from approximately 69\% at 1 degree ($\sim 0.35$~Mpc) to around 5\% at 5 degrees, while the blue fraction also declines but more gradually, from 30\% at 1 degree to 4\% at 5 degrees. Beyond 5 degrees, both red and blue fractions remain low and comparable. 

The trends of the red and blue field dwarfs in Figure \ref{fig:BlueRedWall} are different from those in groups. A major difference is that at all distances from the spine, the fraction of blue field galaxies consistently exceeds that of red ones by roughly 0.2–0.4. For example, at 3 degree, blue galaxies account for about 0.68 while red galaxies are at about 0.32. This agrees with the 2D distribution in Figure \ref{fig:KDE} which showed that red galaxies are sparsely populated outside groups. While it is possible that a part of the blue galaxy population could be due to background or foreground contamination, it is more likely that this difference is an intrinsic distribution as 70\% of the field population are blue galaxies.

The red field galaxy fraction is smaller at 5 deg than at 1 deg similar to group galaxies. However, the slope is not as steep with only a 0.15 drop from 1 deg to 5 deg from the spine. We recall that the field galaxies do not necessarily contain only isolated galaxies. Groups can affect galaxies up to $\sim$ 3 times the virial radius from their centers (\citealt{Cen2014}). Some of the field galaxies may be part of the group or may be infalling galaxies that have been influenced by the group environment (\citealt{Bahe2013}). Previous studies show that infalling galaxies can undergo starbursts that may quench the galaxies before they enter a group (e.g., \citealt{Mahajan2012}). Therefore, the slightly higher fraction of red field galaxies closer to the spine might be due to the galaxies in the outskirts of Fornax cluster, Dorado group and the groups of Eridanus Supergroup. Additional trends of structural and photometric properties (e.g., stellar mass, ellipticity, S\'ersic index, etc.) with distance from the Fornax Wall spine are shown in Appendix \ref{appendix-b}.
\section{Discussion}
\label{sec:discussion}
Matter flows in the Universe from low density regions such as voids to high density regions such as walls and filaments. It is likely that many galaxies within the Fornax Wall originated from nearby environments (R24). The variations in morphology across the different environments in the Fornax Wall may trace the evolutionary history of the supercluster (\citealt{Einasto1980}), similar to how the present structure of galaxies can reflect their dynamical history. 

In the previous section, it was observed that dwarfs in the Fornax Wall are clustered in groups that lie along the spine. As established by T21 and T24, this clustering is mainly observed in red dwarfs while blue dwarfs have a more-or-less uniform distribution throughout the region. They attributed the clustering to known groups like the Fornax cluster. In our work, we observe that the clustering in red dwarfs is mainly found in the large groups situated along the spine of the Fornax Wall. Groups farther from the Fornax Wall do not show significant clustering. This is in agreement with the morphological segregation observed by R24 for massive galaxies, who found that there were more early-type galaxies close to the spine. We also observed that the properties of group galaxies depend on the proximity to the spine. 

Field galaxies show no significant changes in their properties as a function of distance from the spine and $\sim$ 30 \% of field dwarfs were found to be red. For dwarfs, the efficiency of internal feedback is weak (\citealt{Peng2010}) and environmental processes are expected to dominate. These red field galaxies could be objects that quenched their star formation due to early interactions with nearby galaxies, despite residing outside of groups. Low-mass dwarfs are particularly sensitive to such encounters, which can lead to gas removal and suppression of star formation. This is consistent with the findings of \citet{Romero2024}, who show that some low-mass field galaxies have early formation histories.

In the following section, we investigate group and field galaxies by investigating the population of dwarfs that show signs of environmental impact and discuss this with respect to their location in the supercluster.

\subsection{Environmental impact on dwarfs}
\begin{figure*}
    \centering
    \begin{minipage}[b]{0.33\textwidth}
        \centering
        \includegraphics[width = \textwidth]{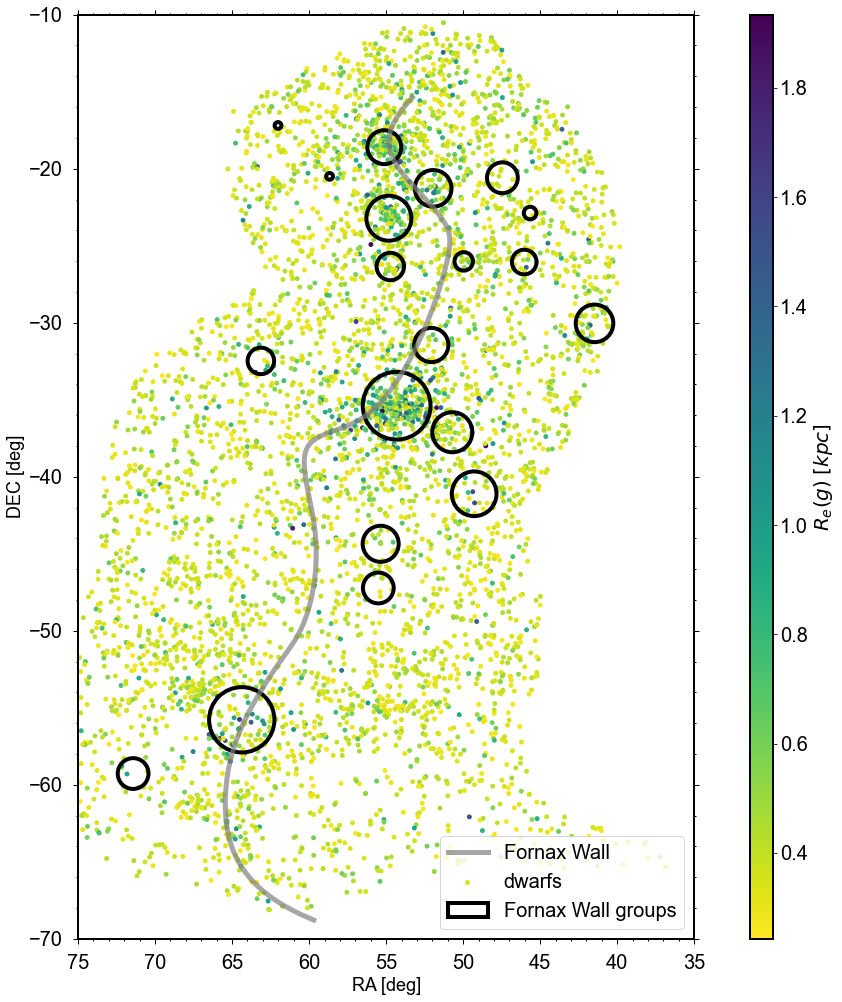}
    \end{minipage}
    \begin{minipage}[b]{0.33\textwidth}
        \centering
        \includegraphics[width = \textwidth]{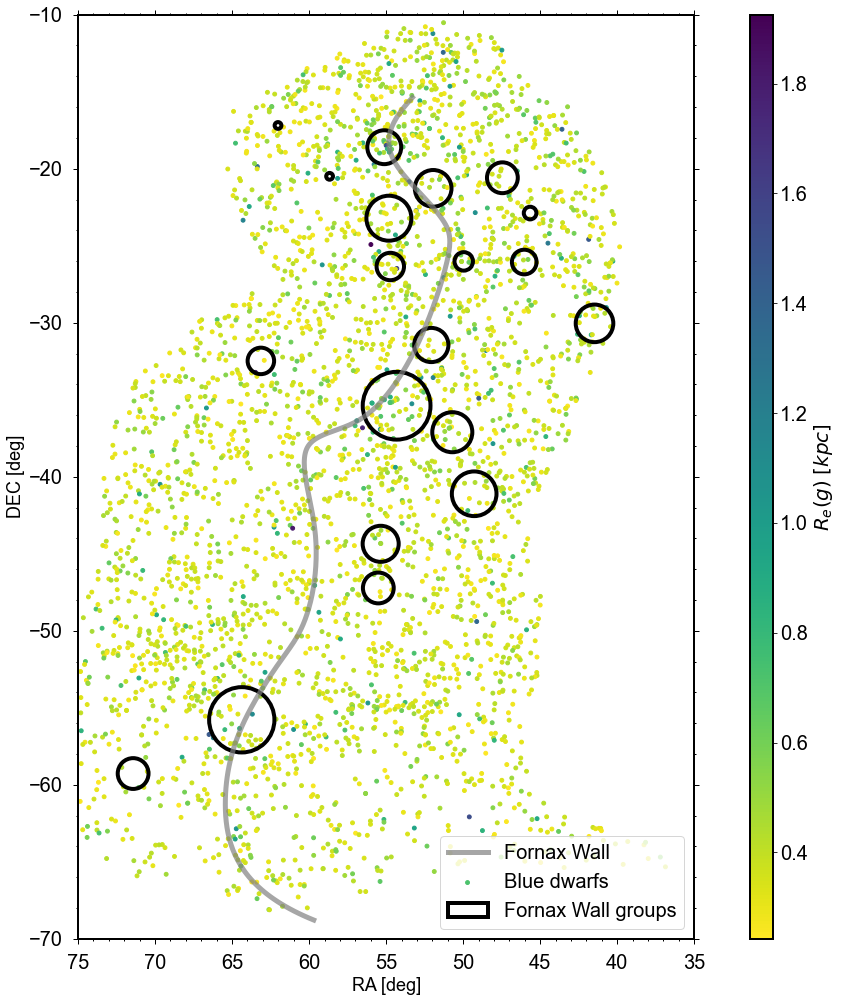}
    \end{minipage}
     \begin{minipage}[b]{0.33\textwidth}
        \centering
        \includegraphics[width = \textwidth]{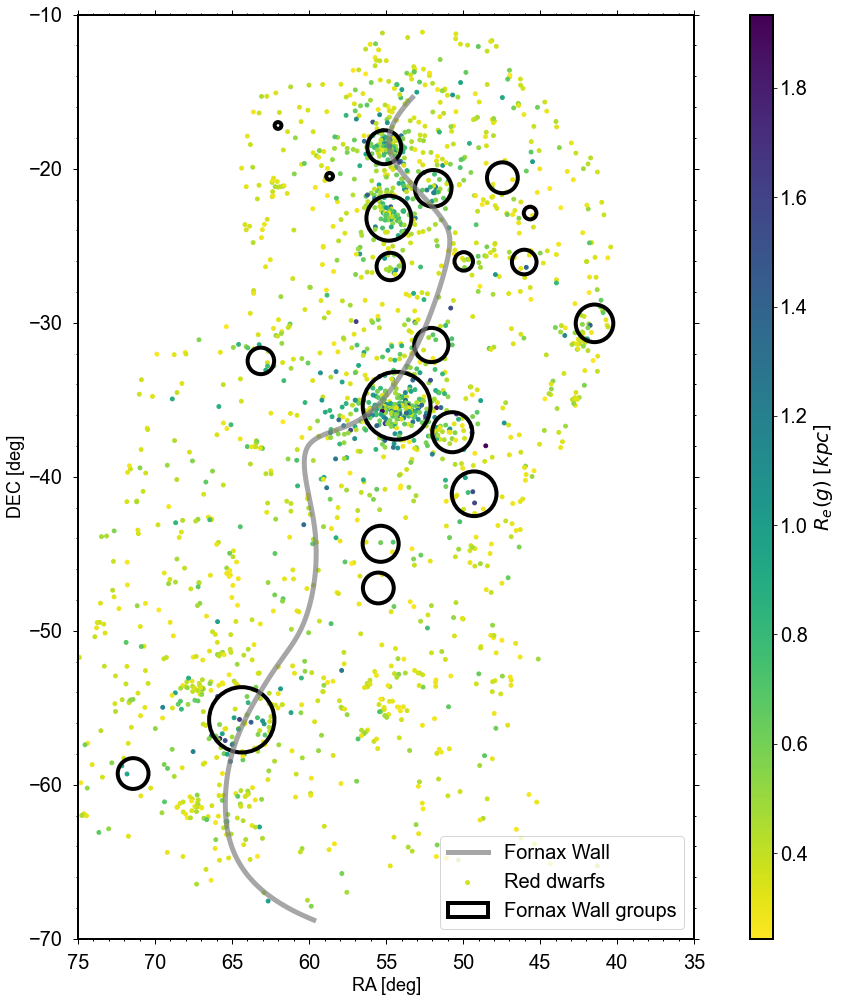}
    \end{minipage}
    \caption{ All the dwarfs (left), blue dwarfs (middle) and red dwarfs (right) in the 2D space with $R_{e,g}$ colourmap. The Fornax Wall spine (gray line) and the Fornax Wall groups (black circles) are shown.
    \label{fig:2Dre}}
\end{figure*}

Although our sample is biased against massive blue dwarfs due to the surface brightness selection threshold (which tends to exclude blue galaxies with high surface brightness at fixed stellar mass), this bias does not affect comparisons between dwarfs of the same colour in different environments. A detailed discussion of this bias is provided in Section \ref{sec:sizedistribution}. We find that the effective radius of galaxies in and around groups is generally higher than in the field environment, as seen in Figure \ref{fig:2Dre}. Figure \ref{fig:2Dre} also shows the distribution of red and blue dwarfs, with the points coloured by $R_{e,g}$. Like with the clustering of galaxies observed by T21 and T24, red galaxies are the main contributors for the difference in $R_{e,g}$ between group and field environments, while blue galaxies have a similar distribution.  

\begin{figure*}
    \begin{minipage}[b]{0.49\textwidth}
        \centering
        \includegraphics[width = \textwidth]{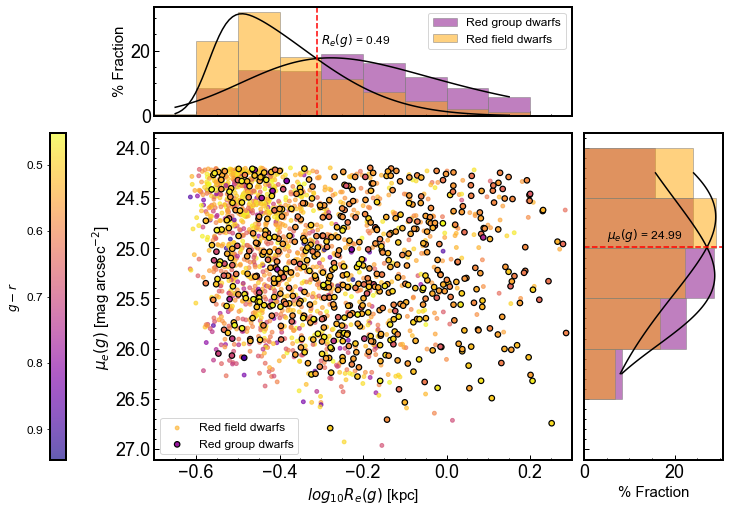}
    \end{minipage}
    \begin{minipage}[b]{0.49\textwidth}
        \centering
        \includegraphics[width = \textwidth]{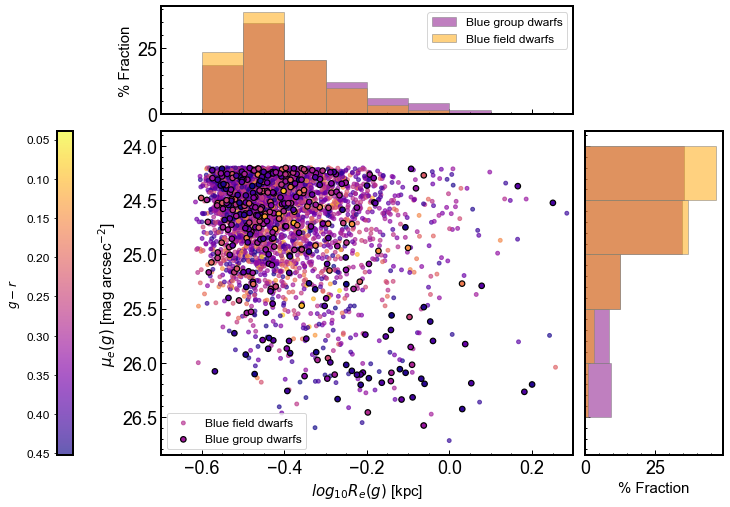}
    \end{minipage}
    \caption{The Kormendy Relation ($R_{e,g}$ vs $\bar{\mu}_{e,g}$) of red (left) and blue (right) dwarfs in the group and field environment with their $g - r$ colour in the colourmap. A skew normal function is fit to the histograms of red dwarfs. The red dashed line corresponds to the intersection point of the red field and group histogram functions.}
    \label{fig:Kormendy}
\end{figure*}

Figure \ref{fig:Kormendy} shows one of the Kormendy relation for the group and field dwarfs in red and in blue. The histograms of the effective radius and effective surface brightness are fit with a skew normal distribution. In the case of effective radius, the two histograms intersect at 0.49 kpc (left panel, for red dwarfs), beyond which the fraction of red field galaxies goes down to zero much faster than the fraction of red group galaxies. When examining blue dwarfs, we find that their distributions in groups and the field environments are very similar, indicating that the differences between group and field populations are significant only for red dwarfs.

\begin{figure}[htbp]
    \centering
    \begin{minipage}{0.5\textwidth}
        \centering
        \includegraphics[width=\textwidth]{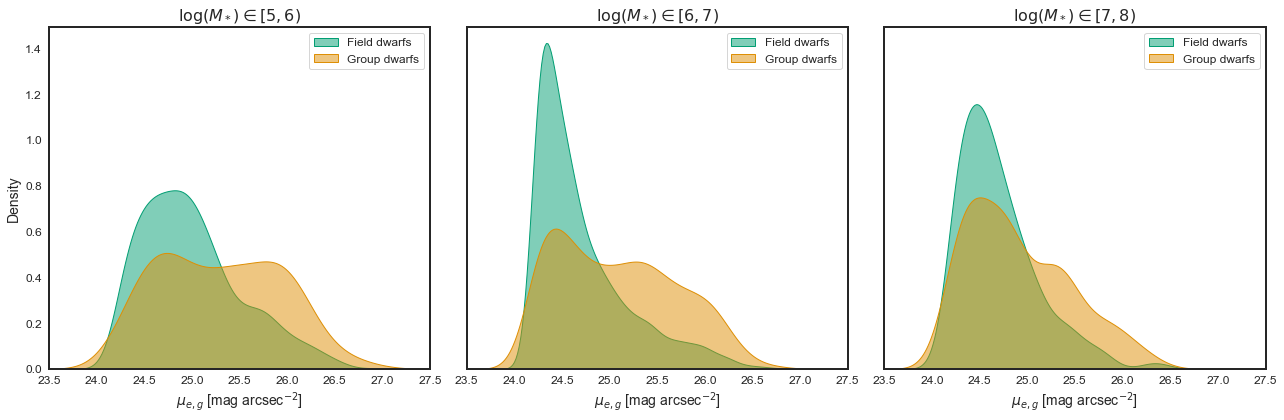}
    \end{minipage}
    \begin{minipage}{0.5\textwidth}
        \centering
        \includegraphics[width=\textwidth]{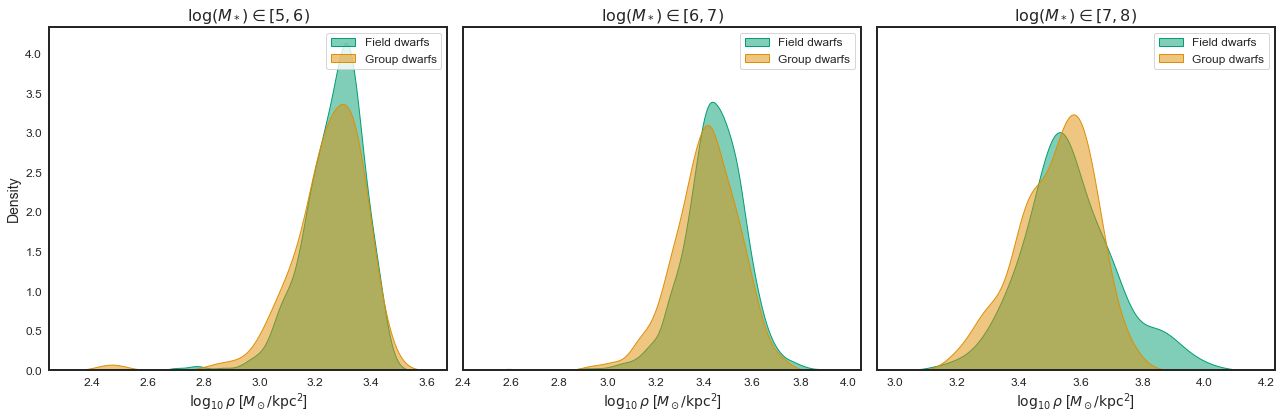}
    \end{minipage}
    \caption{Kernel density distributions of (top) the mean effective surface brightness 
$\mu_{e,g}$ and (bottom) the stellar mass surface density $\rho$ (derived from 
color-based $M/L$ corrections) for field dwarfs (green) and group dwarfs (orange), 
shown in three stellar mass bins.}
    \label{fig:mu_density}
\end{figure}

The comparison of mean surface brightness 
($\mu_{e,g}$) and stellar mass surface density ($\rho$) distributions provides a better understanding of the connection between field and group dwarf populations. Figure~\ref{fig:mu_density}
shows the KDE distributions of $\mu_{e,g}$, where group dwarfs tend 
to have systematically lower surface brightnesses compared to their 
field counterparts. To remove the bias from the 
bluer colors of field dwarfs, we also compute the stellar mass surface 
density $\rho$ using color-based $M/L$ corrections. As shown in the lower panel of
Figure~\ref{fig:mu_density}, the 
difference between group dwarfs and field dwarfs is largely removed. This suggests that the populations can be linked 
through simple fading: once star formation is quenched, field dwarfs evolve to match the properties of group dwarfs, consistent with that environmental quenching primarily affects colours and luminosities rather than 
structural parameters. However, fading alone cannot explain all observed differences. Group dwarfs 
still include a population of objects with larger effective radii, which cannot 
be produced simply by making field galaxies fainter. This indicates that 
additional environmental mechanisms, such as tidal heating, harassment, or 
ram-pressure caused puffing up, may contribute to 
the structural evolution. Therefore, while fading plausibly connects the two populations in terms of brightness and density, the size distribution indicates that structural transformations require further investigation. More precise membership selection and distance measurements will be crucial 
to test these interpretations.

\subsection{Size distribution of dwarfs}
\label{sec:sizedistribution}
Splitting the sample by size at $R_{e,g} = 0.49$ kpc, an arbitrary threshold, we find that nearly 66\% of red galaxies in groups have large radii ($R_{e,g} \geq 0.49$ kpc), while only $\sim$ 30\% of field red galaxies reach this size. In particular, very few field red dwarfs have the largest effective radii, with their fraction dropping to nearly zero at the high end of the size distribution. Thus, the sizes of red early-type dwarfs depend significantly on the environment. There is also an excess of large red dwarfs in groups compared to the field.
This distinction is not seen in blue galaxies where both group and field environments have similar distributions for effective radius in Figure \ref{fig:Kormendy}. If there were large blue galaxies with $M_\star \sim 10^7\,M_\odot$ in groups and clusters, they could not have been included in our sample, as for a fixed stellar mass, blue galaxies tend to have higher surface brightness and fall below our selection threshold of $\bar{\mu}_{e,g} > 24.24$ mag arcsec$^{-2}$. On the other hand, previous studies of groups and clusters find that most objects in these environments are red, particularly at higher stellar masses. When analyzing the data, this selection effect needs to be considered in comparisons between blue and red populations. On the other hand, while red galaxies appear to respond to environmental effects—becoming larger in groups—this trend is not observed for blue dwarfs, albeit they have lower stellar masses than their red counterparts. We find that the peaks of the blue group and field galaxy histograms are not different.

\subsection{Spatial distribution of dwarfs}
The morphology-density relation, first established in rich clusters \citep{Dressler1980}, shows a well-defined correlation between local galaxy density and galaxy type: the fraction of elliptical and S0 galaxies increases, while that of spirals decreases with increasing density. \citet{Bingelli1987-Virgo} confirmed this trend in the Virgo Cluster, finding a strong spatial segregation of morphological types, with early-type (E, S0, dE, dS0) galaxies more concentrated toward the cluster center compared to late-type galaxies (spirals and irregulars). This relationship appears to be universal: \citet{Postman1984} found complete agreement between the morphology-density relations in groups and clusters over the density range of approximately 2 to 200 galaxies Mpc$^{-3}$. They noted that significant morphological changes, such as a decrease in spirals and an increase in S0 and elliptical galaxies, occur above densities of about 5 galaxies Mpc$^{-3}$, while below this threshold the morphological fractions remain similar to those in the field.

As morphological segregation in superclusters is concerned, previous works have been done using massive galaxies. \citet{O'Kane2024} found that galaxies residing in filaments tend to have lower star formation rates and earlier-type morphologies compared to those in the field. This trend persisted even after matching stellar mass, suggesting an environmental origin. Similarly, \citet{Einasto2007a} and \citet{Lietzen2012} reported an excess of passive, early-type galaxies in denser regions of superclusters, consistent with accelerated evolution in high-density environments. \citet{Aghanim2024} further showed that star formation quenching is strongest in supercluster cores and filament intersections. Our results provide a dwarf-scale counterpart to these studies, showing that red dwarf galaxies residing in groups - predominantly located near the Fornax Wall spine - are more abundant, and their fraction decreases with increasing distance from the spine.

Studies focusing specifically on dwarf galaxies in supercluster environments have revealed comparable trends. \citet{Mahajan2010,Mahajan2011}, using 24 micron MIPS data from galaxies with SDSS redshifts, showed that in the Coma supercluster, star formation in massive dwarf galaxies is quenched most effectively in high-density cluster cores, while infalling dwarfs often undergo rapid bursts followed by quenching. Our observation that red dwarf galaxies in group environments tend to have larger effective radii suggests that group-scale environmental processes can influence galaxy structure. Furthermore, \citet{Zanatta2024}, using HST/ACS observations of massive clusters in the Shapley supercluster, find that dwarf galaxies in massive clusters have higher nucleation fractions compared to those in group environments, proving the role of dense environments in shaping dwarf galaxy evolution.

T21 find that large ($R_{e,g} > 0.6$ kpc) and faint ($\bar{\mu}_{e,g} > 26$ mag arcsec$^{-2}$) dwarfs within the DES footprint were almost exclusively red, which, as mentioned, could be due to their sample selection that likely biased against extended blue dwarfs. We find that $\sim 86$\% of the dwarfs in this range are red. Since red galaxies are clustered in groups, and the $R_{e,g}$ and $\bar{\mu}_{e,g}$ distributions found in the Fornax Wall groups and the field environment are different, this must be due to the dwarfs evolving as a result of their environment, for example, when falling into groups or clusters.

\begin{figure}[htbp]
    \centering
    \begin{minipage}[b]{0.24\textwidth}
        \centering
        \includegraphics[width = \textwidth]{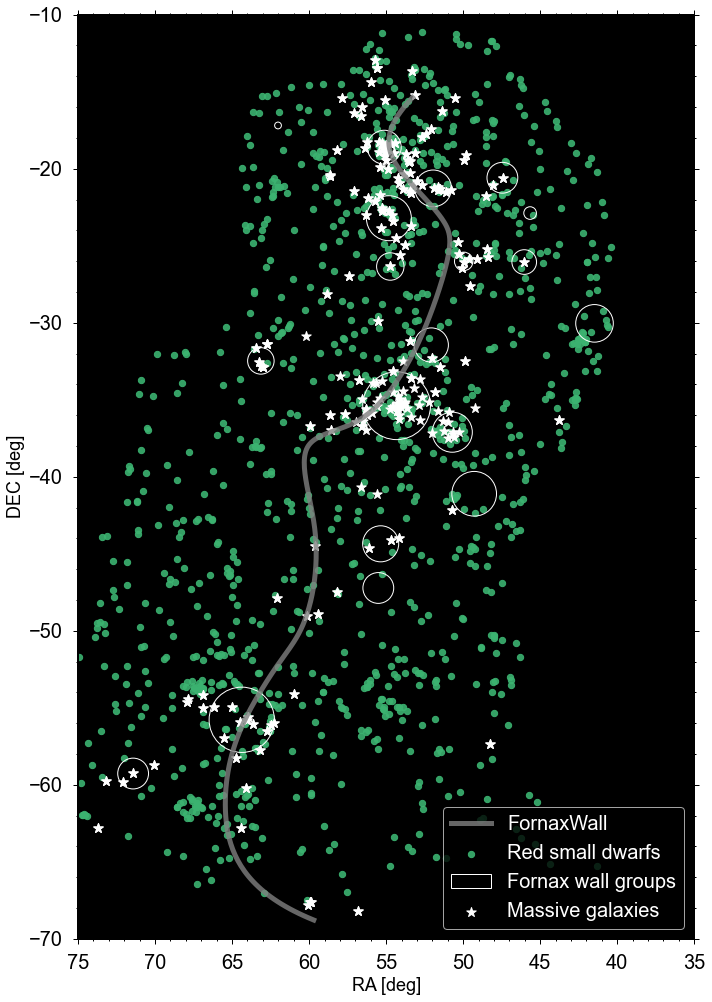}
    \end{minipage}
    \begin{minipage}[b]{0.24\textwidth}
        \centering
        \includegraphics[width = \textwidth]{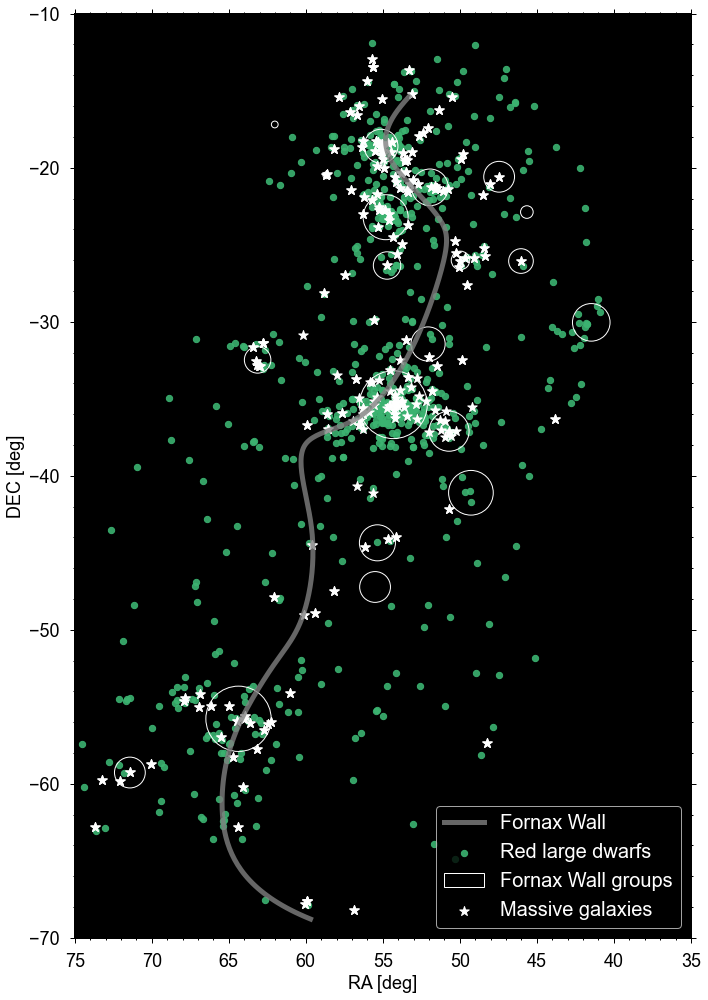}
    \end{minipage}
    \caption{ The 2D distribution of red dwarfs with $R_{e,g} \leq$ 0.49 kpc (left) and red dwarfs with $R_{e,g} >$ 0.49 kpc (right). Massive galaxies (R24) belonging to the Fornax Wall is included as white stars and Fornax Wall groups as white circles. 
    \label{fig:sizeRe2D}}
\end{figure}

Large red dwarfs are found only in groups or next to large groups close to the Fornax Wall in Figure \ref{fig:sizeRe2D}. These regions also have multiple massive galaxies. When compared to the distribution of small dwarfs, they sparsely populate low density regions which suggests that their larger size might be due to interactions in high density regions.

The increase in size of dwarfs in groups may be related to UDG formation. The mechanism that forms the large radius ($R_{e,g} > 1.5$ kpc) of UDGs has been debated in literature. Some suggest that they form through internal mechanisms. For example, in high spin dark matter halos that prevent gas from collapsing into a dense structure (e.g., \citealt{Amorisco2016}), or through outflows due to stellar feedback from starbursts at early epochs (e.g., \citealt{DiCintio2017}). Meanwhile, others suggest that they are formed through environmental influence. For example, failed galaxies that are quenched by ram pressure stripping and tidal interactions before they reach high surface brightness (e.g., \citealt{vanDokkum2015}, \citealt{Martin2019}, \citealt{junais2021}, \citealt{Junais2022}), tidal interactions with massive hosts (e.g., \citealt{Jones2021}), tidal forces in a cluster experienced during/after infall (e.g., \citealt{Sales2020}) or formation of diffuse galaxies from tidal debris created by interacting massive galaxies (e.g., \citealt{Roman2021}). UDGs have been observed both in field (e.g., \citealt{Leisman2017}, \citealt{Roman2019}) and in dense regions (e.g., \citealt{Iodice2020}, \citealt{For2023}). 
Their presence in the field suggests that large, diffuse galaxies can form without environmental influence; however, whether the field UDGs that have been found are exceptions remains unclear due to the lack of complete samples. In our sample, we find 12 UDGs in the field, of which 8 are red and 4 are blue. The existence of red field UDGs may indicate that internal processes play a role in quenching and shaping these galaxies.

In our work, we find that the largest dwarfs are found only in groups close to the Fornax Wall. Many of the formation models listed above predict that diffuse and large galaxies can be found in low density regions. Only a few suggest that they are more likely to be found in groups (e.g., \citealt{vanDokkum2015}, \citealt{Martin2019}). Investigating the formation of large dwarfs in the Fornax Wall may help resolve the formation mechanism of UDGs. 

Based on UDG formation models, \citet{Jackson2021} studied LSBs using a cosmological simulation (NewHorizon). They suggest that LSBs form through a combination of supernova-driven gas outflows, tidal interactions and ram-pressure stripping. In their work, LSBs with lower surface brightness (accompanied by an increase in effective radius) at a given stellar mass are formed from galaxies that were part of high density environment at early epochs. At high redshifts, their progenitors had faster gas accretion which increased star formation and thus, the strength of supernova-driven feedback. The gas outflows due to feedback decreased their surface brightness and increased the effective radius as both gas and old stellar population were pushed out. Both ram pressure stripping and tidal interactions can trigger star formation which then triggers gas outflows. Tidal interactions can also heat up the matter in galaxies, increasing their effective radius. Furthermore, progenitors of LSBs with ``high" surface brightness at a given stellar mass were found in low density environments and did not have strong outflows. Their effective radius grew much slower resulting in smaller galaxies.

Our work supports their theory as nearly all large dwarfs are found in or near groups that are close to the Fornax Wall. However, we do not have direct evidence that their progenitors formed in high density regions at high redshifts as they may have been accreted from nearby regions later (like the UDGs in \citealt{Sales2020}). It is likely that the large dwarfs in the outskirts of groups are infalling galaxies. Therefore, it is still possible that they may have attained their size solely through tidal interactions with massive hosts and/or ram pressure stripping. 

Additionally, our findings are consistent with \citet{Greco2018}, who identified 781 extended LSBGs in $\sim200\,\mathrm{deg}^2$ of imaging data from the Hyper Suprime-Cam Subaru Strategic Program (HSC-SSP). Their sample was divided into red and blue populations: red LSBGs typically show smooth Sérsic profiles, consistent with early-type dwarfs and UDGs commonly found in dense group environments, while blue LSBGs show more irregular morphologies, suggesting ongoing star formation.

In summary, we find that nearly all red, large dwarfs are found within or near the Fornax Wall groups, where the brightest group galaxies and other massive galaxies are located, while small red dwarfs dominate the field environment. In particular, the largest red dwarfs are only found in groups. This likely supports the simulation by \citet{Jackson2021} which suggests that they were formed in high density environments through supernova feedback induced by the environment. Additional investigation is necessary to understand if large red dwarfs are mainly influenced by the local environment (group and massive host galaxies) or if the large-scale assembly of galaxies and matter (infall or increased gas accretion at early epochs) is also important. 

We also comment on the distribution of small red dwarfs, as the presence of quenched dwarfs in the field environment could present an interesting avenue for future research. In some regions, they seem to be clumped together (red circles in Figure \ref{fig:sizeRe2D}). These regions do not always have a massive galaxy in them (e.g., below the Dorado group, towards the left in the figure) and are not necessarily near known groups. We speculate that they may be dwarf-galaxy groups with no massive hosts like those observed by \citet{Tully2006}. Further analysis is required to confirm if they form groups, if their star formation is quenched and to estimate if they may be precursors of pre-processed cluster galaxies. 

   \section{Summary and conclusions}
   \label{sec:conclusion}
In this work, we explored the dwarf galaxies in the Fornax-Eridanus supercluster using the spine of the Fornax Wall defined by R24. We obtained a supercluster low-surface brightness dwarf galaxy (SC dwarf) catalog by identifying Fornax Wall member galaxies from the DES dwarf catalog using scaling relations, a known background/foreground group catalog and visual inspection. We calculated the absolute magnitude and stellar masses of the galaxies, and their distributions were in good agreement with literature. The SC dwarf catalog was estimated to contain mostly dEs and dIrrs based on the size-magnitude relation. We summarise the main results: 

\begin{itemize}
    \item [1. ]  We investigated the clustering of red dwarfs found in the DES survey by T21 and T24, in the context of the supercluster. Red dwarfs ($g - r < 0.45$) dominated the group environment while blue dwarfs dominated the field environment as determined by T21 and T24 
    \item [2. ]  We find that the morphology of group dwarfs depends on their proximity to the Fornax Wall spine. Red dwarfs were located close to the spine and their fraction decreased away from the spine. 
    \item [3. ] The effective radius of red dwarfs was found to depend on the environment. The largest red dwarfs are found in groups, and much fewer are found in the field environment.
\end{itemize}

The morphological segregation of massive galaxies in the Fornax Wall was attributed to pre-processing in groups and the inflow of galaxies from nearby structures by R24. Our work suggests that dwarfs also follow a similar morphological segregation with more red dwarfs found close to the spine indicating that pre-processing may play a role. In addition to this, we find that the environment in groups is likely to increase the effective radius of red dwarfs. As the largest red dwarfs are not found in the field environment, it is likely that the group environment plays a primary role in their formation.

\subsection{Suggestions for future work}

The SC dwarf catalog, while comprehensive, includes galaxies that are not dwarfs, such as bright galaxies with S\'ersic $n \sim 5$, and those with $\bar{\mu}_{e,r} > 23.5$ mag arcsec$^{-2}$. In addition to this, both T21 and T24 catalogues, from which the SC dwarf catalog is derived, were dominated by blue galaxies. To enhance the accuracy of the galaxy properties and refine the purity of the dwarf sample, it is crucial to assess whether the objects in T21 and T24 are genuine dwarfs and determine their distances.

To test the completeness of the T21 catalogue, we compared it with the FDSDC catalogue, for which membership classification was performed in great detail, in the Fornax Cluster. We find that, in the stellar mass range of interest, approximately 48\% of the objects were found. However, in the same area, about 17\% of the objects are not in FDSDC and are most likely false positives. In the field, it is much harder to verify membership, but we estimate that the fraction of false positives is much higher.

The Euclid (\citealt{Laureijs2011}) mission surveys $\sim$ 15000 deg$^2$ area with high spatial resolution and depth that has not been achieved previously for wide surveys (\citealt{Eucliddepth}). Imaging from the Visual instrument (VIS; \citealt{Cropper2014}) in the optical band is able to resolve features in dwarfs and may enable the use of surface brightness fluctuations to estimate distances to these galaxies (\citealt{Cuillandre2024}). In addition, the improved resolution will make visual identification of dwarf galaxies significantly easier. The advent of Euclid and future wide surveys with deep data will largely improve the identification of members of the Fornax Wall. 

Analysing additional properties such as metallicity, asymmetry, and alignment of the galaxies can give deeper insights into their star formation histories and structure. \citet{Venhola2021} showed that dwarf galaxies with large excess size (the deviation of the effective radius of a galaxy from the mean at a given mass) had a weak preferential alignment towards the Fornax cluster center. Moreover, galaxies that have experienced tidal disturbances may be elongated towards the perturbation source. Investigating this can reveal whether the Fornax wall may have some (weak) influence on field dwarfs and confirm whether pre-processing dictates the observed morphological distribution.

Applying a nearest-neighbor algorithm to detect possible groups of dwarfs in the field environment can provide better insights on the mechanisms responsible for the properties of galaxies in them. Analysing filamentary structures detected by R24 around the Fornax Wall is also vital to understand the evolutionary pathway of galaxies and the subsequent environmental effects on dwarfs, especially in the field environment. To conclude, our understanding of how environment influences dwarf galaxy evolution is still in its early stages.    

\begin{acknowledgements}
J. is funded by the European Union (MSCA. EDUCADO, GA 101119830 and WIDERA ExGal-Twin, GA 101158446)
\end{acknowledgements}

\bibliographystyle{aa} 
\bibliography{references} 

\begin{appendix} 
\onecolumn
\section{Comparison of T21 and FDSDC Catalogues} \label{appendix-a}
\begin{figure*}
    \centering
    \begin{minipage}[b]{0.4\textwidth}
        \centering
        \includegraphics[width = \textwidth]{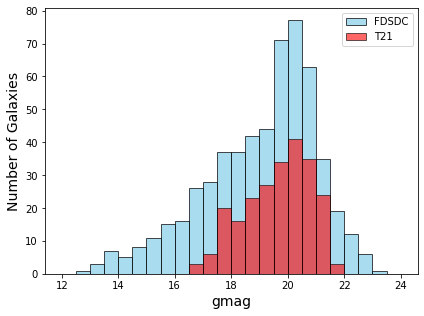}
    \end{minipage}
    \begin{minipage}[b]{0.4\textwidth}
        \centering
        \includegraphics[width = \textwidth]{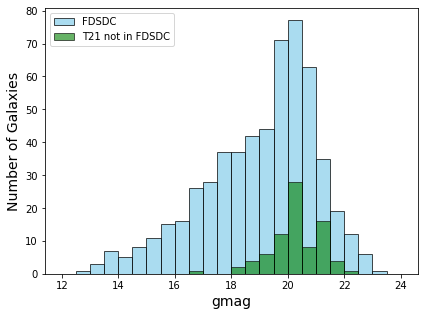}
    \end{minipage}
        \begin{minipage}[b]{0.4\textwidth}
        \centering
        \includegraphics[width = \textwidth]{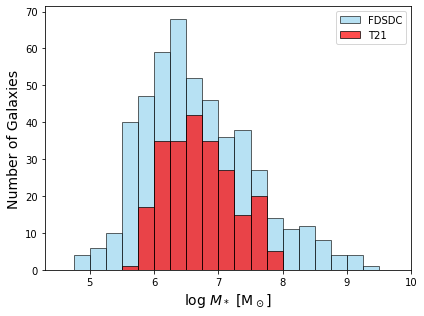}
    \end{minipage}
        \begin{minipage}[b]{0.4\textwidth}
        \centering
        \includegraphics[width = \textwidth]{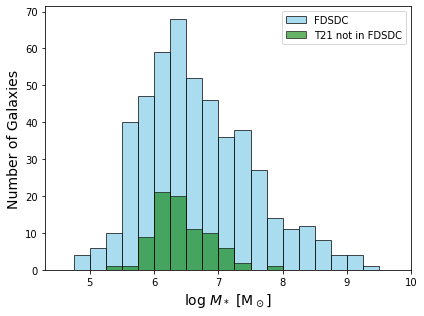}
    \end{minipage}
    \caption{Comparison between the T21 and FDSDC catalogues in the Fornax Cluster. Top-left: $g$-band magnitude distribution of FDSDC sources (blue) and the T21–FDSDC crossmatched sources (red). Top-right: $g$-band magnitude distribution of FDSDC sources (blue) and T21 sources not present in FDSDC (green). Bottom-left: stellar mass distribution for the matched sample (red) and FDSDC sample (blue). Bottom-right: stellar mass distribution of unmatched sample (green) and FDSDC sample (blue).
    \label{fig:t21_v18_comp}
}
\end{figure*}

To better investigate the completeness of the T21 catalogue relative to the FDSDC catalogue, in Figure \ref{fig:t21_v18_comp} we show $g$-band magnitude and stellar mass distributions of all FDSDC dwarf galaxies (blue), those that are matched with the T21 catalogue (red), and T21 galaxies that are not found in the FDSDC catalogue (green).

\section{Structural parameters of the dwarfs} \label{appendix-b}
\begin{figure*}
    \begin{minipage}[l]{0.65\linewidth}
        \includegraphics[width=\textwidth]{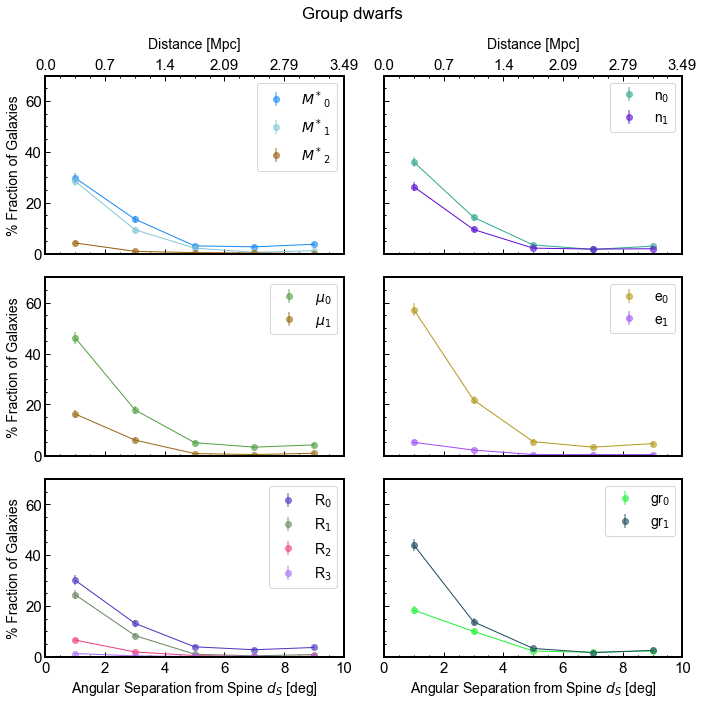} 
    \end{minipage} 
    \begin{minipage}[b]{0.2\linewidth}
        \begin{tabular}{>{\footnotesize}c >{\footnotesize}c}
            \hline\hline
            \textbf{Label} & \textbf{Bin}\\
            \hline
            $\mu_0$ & $\bar{\mu}_{e, r} < 25$ mag arcsec$^{-2}$ \\
            $\mu_1$ & $\bar{\mu}_{e, r} \geq 25$ mag arcsec$^{-2}$\\
            \hline
            $M^*_0$ & $\log \; M^* \; < 6.5 \; M_\odot$ \\
            $M^*_1$ & $6.5 \leq \log \; M^* \; < 7.5 \; M_\odot$ \\
            $M^*_2$ & $ \log \; M^* \geq 7.5 \; M_\odot$ \\
            \hline
            n$_0$ & $\text{S\'ersic } n \leq 1$\\
            n$_1$ & $\text{S\'ersic } n > 1$ \\
            \hline
            R$_0$ & $R_{e, r} \; < 0.5 \; \text{kpc}$  \\
            R$_1$ & $0.5 \leq R_{e, r} \; < 1 \; \text{kpc}$ \\
            R$_2$ & $1 \leq R_{e, r} \; < 1.5 \; \text{kpc}$ \\
            R$_3$ & $ R_{e, r} \geq 1.5 \; \text{kpc}$ \\
            \hline
            e$_0$ & $\text{ellipticity} < 0.5$ \\
            e$_1$ & $\text{ellipticity} \geq 0.5 $ \\
            \hline
            gr$_0$ & $g - r < 0.452$ \\
            gr$_1$ & $g - r \geq 0.452$ \\
            \hline
        \end{tabular}
    \end{minipage}
    \caption{Fraction of group dwarfs as a function of projected distance from the Fornax Wall spine for $M_*$, S\'ersic n, $\bar{\mu}_{e,r}$, ellipticity, $R_{e,r}$, and $g - r$ colour bins (left). Table with the labels in Column 1 and the corresponding bins in Column 2 (right).\label{fig:GroupWall}}
\end{figure*}

In the previous section, we established that groups near the spine have more red galaxies than ones farther. We investigate if other galaxy properties show such a relation. The stellar mass $M_*$, effective surface brightness $\bar{\mu}_{e,r}$, effective radius $R_{e,r}$, S\'ersic index $n$ and the ellipticity have been divided into different bins provided in the table in Figure \ref{fig:GroupWall}. The effective radius bin \textbf{($R_3$)} contains the UDG candidates identified in Section \ref{sec:SizeMag}. The two S\'ersic index bins each define flat and bulgy galaxies. The $g - r$ bins retain the definitions in Section \ref{sec:colourcolour} for red and blue galaxies. The remaining bins have been defined by dividing the sample range into equal sizes. 

The fraction of group dwarfs for these bins is shown in Figure \ref{fig:GroupWall}.  Within 1 deg from spine, there are 55\% more group galaxies with log $M_* < 7.5 \; M_\odot$, 30\% more galaxies with red colour, 55\% more galaxies with ellipticity $< 0.5$, 35\% more galaxies with $\bar{\mu}_{e,r} < 25$ mag arcsec$^{-2}$ and 35\% more galaxies $R_{e,r} < 1$ kpc as compared to galaxies outside these ranges. Regardless of whether the population in the different bins mentioned above are the same galaxies or not, we speculate that there is some effect of the supercluster structure on group galaxies depending on their proximity to the spine.  

\begin{figure*}
    \begin{minipage}[l]{0.65\linewidth}
        \includegraphics[width=\textwidth]{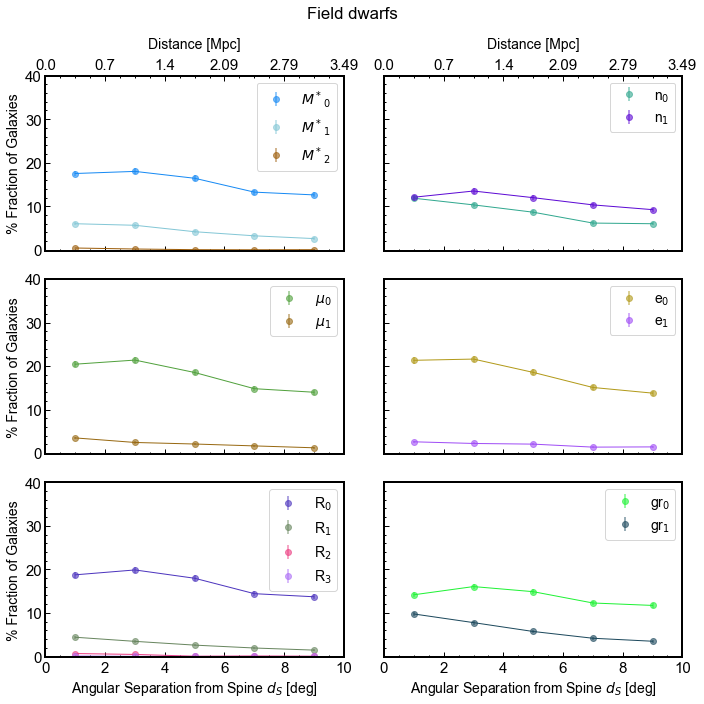} 
    \end{minipage} 
    \begin{minipage}[b]{0.2\linewidth}
        \begin{tabular}{>{\footnotesize}c >{\footnotesize}c}
            \hline\hline
            \textbf{Label} & \textbf{Bin}\\
            \hline
            $\mu_0$ & $\bar{\mu}_{e, r} < 25$ mag arcsec$^{-2}$ \\
            $\mu_1$ & $\bar{\mu}_{e, r} \geq 25$ mag arcsec$^{-2}$\\
            \hline
            $M^*_0$ & $\log \; M^* \; < 6.5 \; M_\odot$ \\
            $M^*_1$ & $6.5 \leq \log \; M^* \; < 7.5 \; M_\odot$ \\
            $M^*_2$ & $ \log \; M^* \geq 7.5 \; M_\odot$ \\
            \hline
            n$_0$ & $\text{S\'ersic } n \leq 1$\\
            n$_1$ & $\text{S\'ersic } n > 1$ \\
            \hline
            R$_0$ & $R_{e, r} \; < 0.5 \; \text{kpc}$  \\
            R$_1$ & $0.5 \leq R_{e, r} \; < 1 \; \text{kpc}$ \\
            R$_2$ & $1 \leq R_{e, r} \; < 1.5 \; \text{kpc}$ \\
            R$_3$ & $ R_{e, r} \geq 1.5 \; \text{kpc}$ \\
            \hline
            e$_0$ & $\text{ellipticity} < 0.5$ \\
            e$_1$ & $\text{ellipticity} \geq 0.5 $ \\
            \hline
            gr$_0$ & $g - r < 0.452$ \\
            gr$_1$ & $g - r \geq 0.452$ \\
            \hline
        \end{tabular}
    \end{minipage}
    \caption{Fraction of field dwarfs as a function of projected distance from the Fornax Wall spine for $M_*$, S\'ersic n, $\bar{\mu}_{e,r}$, ellipticity, $R_{e,r}$, and $g - r$ colour bins (left). Table with the labels in Column 1 and the corresponding bins in Column 2 (right). The table is the same as in Figure \ref{fig:GroupWall}\label{fig:PristWall}}
\end{figure*}

The fraction of field dwarfs is shown in Figure \ref{fig:PristWall}. All the properties have nearly constant fractions in every bin. Any difference in the fractions between each distance bin is $<5$\% for all the properties except for ellipticity, for which the difference is $<10$\%. These trends are similar to those observed for $g - r$ colour and suggests that there may be no environmental effect on field galaxies as a function of distance to the spine. We discuss the distribution of these properties in the next section. 

\section{Quantifying Projection Contamination Using Simulations} \label{appendix-c}
One of the main limitations in studying faint dwarf galaxies is the lack of direct distance measurements, which makes their environmental classification uncertain. In this work, we therefore assume that dwarf galaxies projected near massive galaxies and groups in the Fornax-Eridanus region are physically associated with the same large-scale structures. To test the validity of this assumption, we perform an independent validation using a cosmological hydrodynamical simulation.

\subsection{Simulation Data}
We use the high-resolution 25 Mpc/h cosmological volume from the SIMBA simulation \citep{Dave2019}. All dark matter halos with masses $M > 10^{10} M_\odot$ are extracted from the simulation. Stellar masses are assigned to these halos using DarkLight, a semi-empirical dwarf galaxy formation model designed to robustly predict the stellar mass–halo mass relation for low-mass galaxies \citep{Kim2024}. 
This results in a complete sample of 4,957 dwarf galaxies with stellar masses in the range $10^{6} \leq M_* \leq 10^{9} M_\odot$, matching the mass range of our observed sample. Massive group-central galaxies are defined as systems with $M_* > 10^{11} M_\odot$, corresponding to halos with mean masses of several $10^{12} M_\odot$. There are 31 such group-central galaxies in the simulated volume. Because the full three-dimensional positions and halo memberships of all galaxies are known, the simulation provides a ground-truth environmental classification, allowing a direct comparison with projected (observational-like) group and filament assignments.

\subsection{Filament Identification}
Filaments are identified using the 1-DREAM algorithm \citep{Canducci2022}.  First, we apply the LAAT tool to highlight filamentary regions. The main parameters are: 7³ ants, a neighbourhood radius of 3 Mpc, and a pheromone threshold that removes the lowest 25\% of values, which enhances the contrast of filamentary features. We checked that reasonable changes to these parameters do not affect our results.
Next, we use the MBMS tool to extract the filament skeleton, using the same 3 Mpc neighbourhood radius as in the LAAT step. Finally, we use the DimIndex tool to select the one-dimensional structures from the MBMS points, again with a neighbourhood radius matching that of LAAT. These points form the final skeleton of filaments in the volume. Galaxies within 3 Mpc of the filament skeleton are classified as filament members. We additionally test a more conservative threshold of 2 Mpc. Because of the well-known tendency of massive halos to reside preferentially near filaments, all group galaxies in the volume (31 groups in total) are naturally identified as filament members in this procedure.

\subsection{Projection Tests}
\begin{figure}[htbp]
    \centering
    \includegraphics[width=0.6\columnwidth]{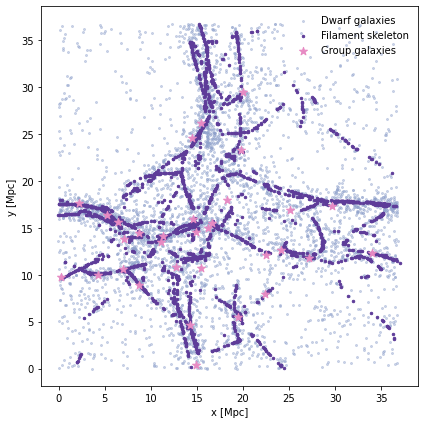}
    \caption{\label{fig:projection} Two-dimensional projection of the SIMBA simulation volume used to quantify projection effects. Grey points show dwarf galaxies ($10^6 \leq M_* \leq 10^9 M_\odot$), pink stars indicate massive group galaxies ($M_* > 10^{11} M_\odot$), and purple points show the filament skeleton identified using the 1-DREAM algorithm.}
\end{figure}

To assess projection effects, we project the simulated volume along three independent lines of sight (x, y, and z). For each projection, we select dwarf galaxies appearing within one virial radius ($R_{200}$) of a group central (using the same selection criteria as in the observations) and measure the fraction that are true filament members in 3D. 
The fractions of dwarfs genuinely associated with filaments are 98.9\%, 96.2\%, and 94.6\% for the x, y, and z projections, respectively, giving a combined fraction of 96.6\%. Using a stricter filament definition (within 2 Mpc of the skeleton), the corresponding fractions are 92.4\%, 87.0\%, and 82.8\%, with a combined value of 87.4\%. We also quantify line-of-sight contamination: $\sim$20\% of dwarfs projected within $R_{200}$ are actually >10 Mpc from the group, but 98\% of these are still filament members. A visual illustration of the spatial relationship between dwarf galaxies, groups, and filaments in the simulation is shown in Figure \ref{fig:projection}.
These results show that dwarfs projected near group centrals are overwhelmingly associated with the cosmic web. This supports the assumption adopted in the main text that dwarfs near massive galaxies trace the same large-scale structures, and that statistical environmental trends derived from projected samples are robust. Projection effects do not significantly bias the environmental trends reported in this work.

\section{Completeness of the DES catalogue from mock-galaxy injection tests} \label{appendix-d}
To estimate the completeness of the DES-based catalogues used in this work, we performed mock galaxy injection and recovery tests on random DES tiles. We generated mock dwarf galaxies using the GalSim package \citep{Rowe2014}. Each mock galaxy was modelled as a single-component Sérsic profile, convolved with the DES point spread function, and Poisson noise was added before injection into the DES multiband images. The injected galaxies were placed at random positions on the tiles, with the requirement that no two mock dwarfs overlap. Their structural and photometric parameters — including size, magnitude, colour, and Sérsic index — were randomly drawn from the same ranges spanned by the DES low-surface-brightness galaxy samples of T21 and T24.

For each realization, 100 mock dwarfs were injected into a DES tile, and this procedure was repeated 100 times, resulting in a total of 10,000 injected mock galaxies. After injection, we ran the same SourceExtractor configuration used to build the DES catalogues (\citealt{Morganson2018}, Table~24) and recorded whether the injected sources were recovered by the detection pipeline.

\begin{figure*}
    \centering
    \begin{minipage}[b]{0.4\textwidth}
        \centering
        \includegraphics[width = \textwidth]{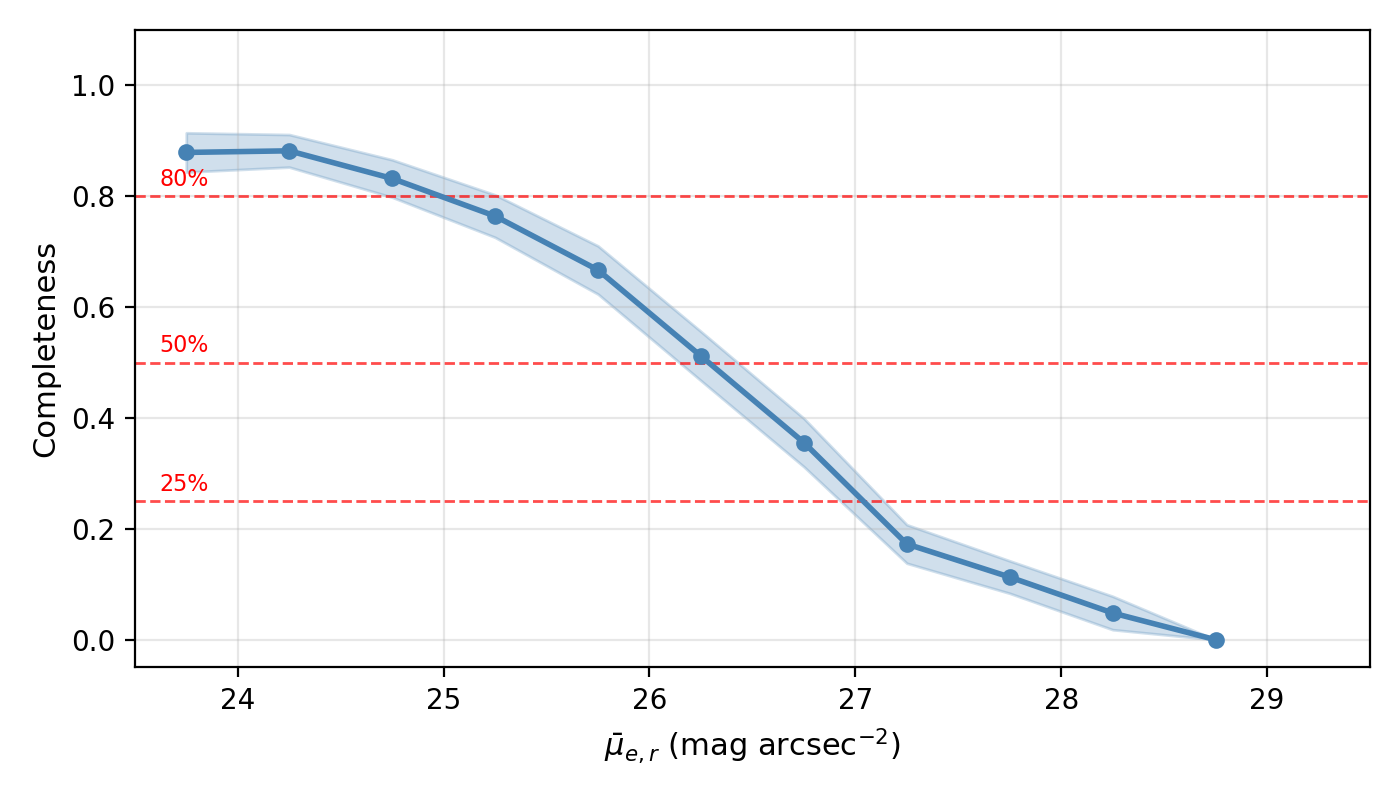}
    \end{minipage}
    \begin{minipage}[b]{0.4\textwidth}
        \centering
        \includegraphics[width = \textwidth]{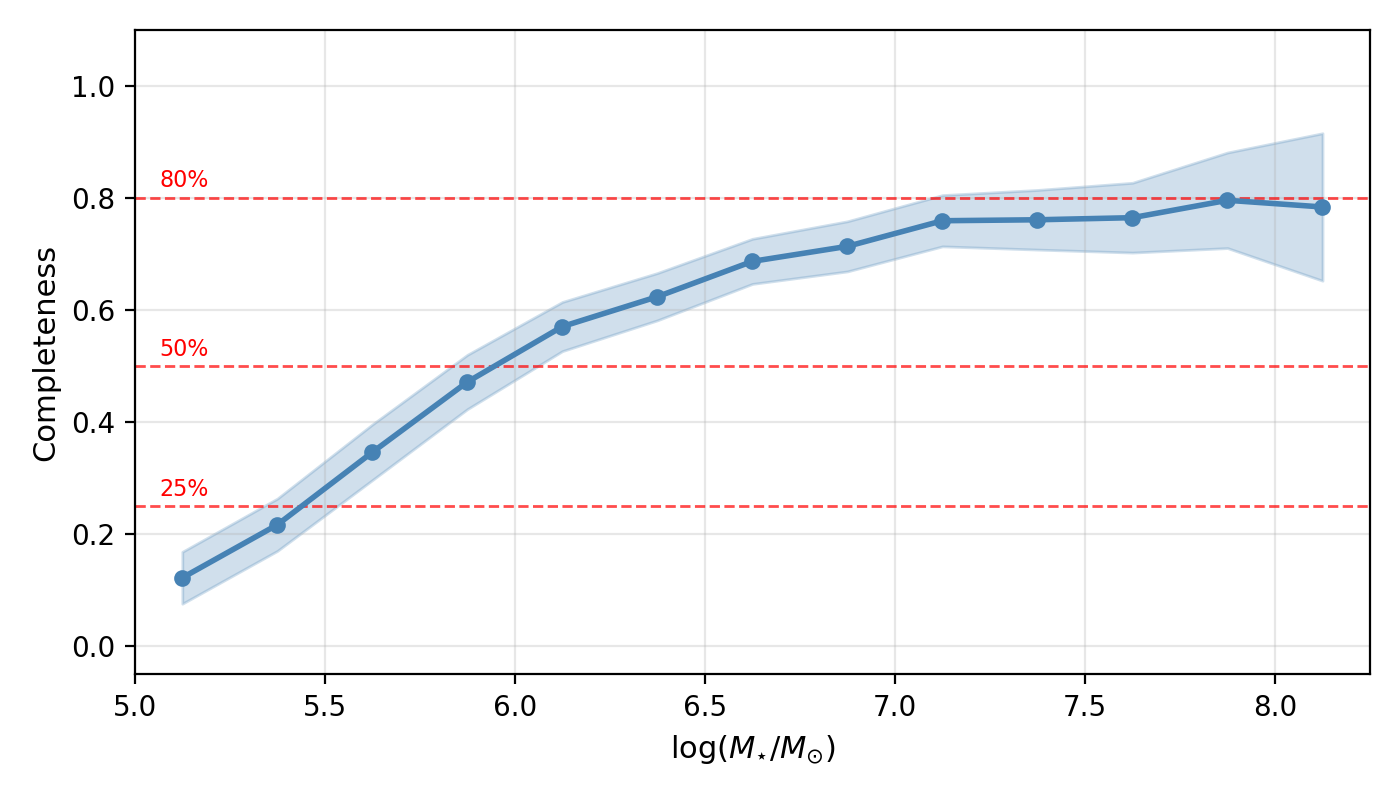}
    \end{minipage}
    \caption{Completeness of the injected mock dwarf galaxies as a function of mean effective $r$-band surface brightness (left panel) and stellar mass (right panel). The shaded region shows the scatter among realizations.
    \label{fig:comp_curve}
}
\end{figure*}

The resulting recovery fractions are shown in Figures \ref{fig:comp_curve} as a function of mean effective r-band surface brightness and stellar mass, respectively. To estimate stellar masses for the injected galaxies, we assumed a distance of 20 Mpc and applied the colour-based stellar mass-to-light relation of \citet{Taylor2011b} using the injected $g-i$ colour.

These tests show that the DES parent catalogue remains reasonably complete at relatively high surface brightness, but the completeness declines significantly toward fainter and lower-mass systems. As a function of surface brightness, the recovery fraction is about 80\% at $\bar{\mu}_{e,r} \approx 25\,\mathrm{mag\,arcsec^{-2}}$, decreases to about 50\% at $\bar{\mu}_{e,r} \approx 26\,\mathrm{mag\,arcsec^{-2}}$, and declines rapidly at still fainter levels. As a function of stellar mass, the recovery fraction is about 80\% in the range $7 \lesssim \log(M_\star/M_\odot) \lesssim 8$, while it decreases below $\log(M_\star/M_\odot) \approx 7$ and reaches about 50\% near $\log(M_\star/M_\odot) \approx 6$.

These recovery fractions correspond only to the SourceExtractor detection stage of the DES catalogue. Because the subsequent machine-learning selection, Sérsic fitting, and visual inspection applied in T21 and T24 are not included in these tests, the values reported here should be regarded as upper limits on the completeness of the final dwarf catalogues.

\end{appendix}
\end{document}